# Origin of Reduced Coercive Field in ScAlN: Synergy of Structural Softening and Dynamic Atomic Correlations

[1]Ryotaro Sahashi, [1]Po-Yen Chen, and [1,2]Teruyasu Mizoguchi (corresponding author)
Email: teru@iis.u-tokyo.ac.jp
[1]Department of Materials Engineering, The University of Tokyo, Tokyo, Japan
[2]Institute of Industrial Science, The University of Tokyo, Tokyo, Japan

## Abstract
Among wurtzite-type ferroelectrics, scandium-doped aluminum nitride (ScAlN) has emerged as a leading candidate for CMOS-compatible low-voltage memory, combining strong spontaneous polarization with process compatibility. A remarkable feature of this system is the pronounced reduction of the coercive field ($E_c$) with increasing Sc concentration; however, its microscopic origin remains poorly understood at the atomic scale, particularly under finite temperature and applied electric fields. Here, we employ an atomistic simulation framework that integrates a density-functional-theory-accurate machine-learning force field with an equivariant neural-network-based Born effective charge model to perform large-scale electric-field-induced molecular dynamics simulations at near-first-principles accuracy. The framework correctly reproduces key experimentally observed trends, including the decrease in the c/a ratio and the monotonic reduction of $E_c$ with increasing Sc content. Beyond static structural softening, we uncover a dynamic mechanism underlying $E_c$ reduction. Sc atoms exhibit larger thermal vibrations and begin to displace earlier during switching, acting as dynamic triggers for polarization reversal. Moreover, the displacement correlation between Sc and Al atoms evolves systematically with composition, enhancing cooperative atomic rearrangements and lowering the effective switching barrier. These results demonstrate that $E_c$ reduction in ScAlN arises from the synergy of structural softening and dynamic correlation evolution, providing a new perspective for designing hexagonal ferroelectrics.

## Introduction
Ferroelectric random-access memory (FeRAM) has attracted considerable attention as a key device for realizing in-memory computing (IMC) in emerging artificial intelligence (AI) hardware, owing to its ability to achieve both fast write speeds and low power consumption.[1–3] Among candidate materials, scandium-doped aluminum nitride (ScAlN) has garnered strong interest from both industry and academia because, unlike conventional perovskite oxides, it combines excellent ferroelectric properties with high compatibility with CMOS fabrication processes.[4–8]

A particularly notable feature of ScAlN is the dramatic reduction of the coercive field ($E_c$) with increasing Sc concentration, which opens the path toward low-voltage device operation.[9–11] Nevertheless, the $E_c$ values remain substantial, posing a critical bottleneck for practical low-voltage operations in logic-in-memory architectures. While the macroscopic reduction of $E_c$ is well-established, the lack of a comprehensive understanding of the underlying atomistic switching dynamics hinders the rational design of next-generation ferroelectrics with further optimized performance. To date, the mechanism of $E_c$ reduction has been primarily discussed based on static energy analyses using first-principles density functional theory (DFT) calculations.[12] Previous studies have pointed out that the incorporation of Sc relaxes the structural anisotropy of the crystal lattice and reduces the energy difference between the polar wurtzite phase and the nonpolar hexagonal phase. This so-called "structural softening" lowers the polarization reversal barrier on the static potential energy surface (PES).[13]

In contrast, actual ferroelectric switching is an inherently dynamic process that proceeds under finite temperature and applied electric fields. Static PES analyses alone are insufficient to fully capture the dynamic picture, such as how atoms overcome the energy barrier through thermal fluctuations or how Al and Sc atoms cooperate via distinct roles and correlated responses during polarization reversal. To address these dynamic aspects, ab initio molecular dynamics (AIMD) based on DFT has been widely employed to investigate electric-field-induced structural changes.[14,15] However, due to its extremely high computational cost, AIMD is restricted to limited spatial and temporal scales, making it difficult to achieve the statistical accuracy required to discuss composition dependence and correlated atomic motions in solid-solution systems.

To overcome these limitations, computational approaches based on empirical simulations and machine-learning force fields (MLFFs) have increasingly been applied in materials science in recent years.[16–19] In particular, MLFFs have advanced rapidly through the development of graph neural network-based frameworks such as MACE, ORB, and SevenNet,[20–22] as well as descriptor-based approaches such as GRACE. Indeed, MLFF-based molecular dynamics (MD) simulations have been shown to accurately reproduce structural and dynamical properties consistent with first-principles calculations across a wide range of environments—including crystalline, liquid, and amorphous phases—in prototypical ferroelectrics such as $BaTiO_3$, thereby demonstrating their reliability for large-scale and long-time MD simulations.[23] MLFFs are data-driven models trained on high-accuracy first-principles datasets, enabling the prediction of interatomic forces and energies with near-DFT accuracy. Furthermore, approaches that integrate ML potentials with Born effective charge (BEC) prediction models have been successfully applied to representative ferroelectrics such as $BaTiO_3$ and $HfO_2$.[24,25] Nevertheless, these studies have been largely limited to perovskite or fluorite structures, and studies applying these approaches to wurtzite-structured ScAlN to elucidate its switching mechanism in detail remain scarce.

In this study, we aim to physically elucidate the origin of the reduced $E_c$ in ScAlN by integratively analyzing both “static structural softening” and “dynamic atomic motion.” To this end, we utilize a fine-tuned MLFF based on MACE,[26] which retains DFT-level accuracy while enabling large-scale and long-time simulations, and integrate it with an equivariant neural network-based BEC prediction model.[27] This framework allows us to atomistically track large-scale switching processes under applied electric fields that are inaccessible to conventional static calculations or AIMD approaches. [24,25] We demonstrate that, in addition to the static lattice destabilization induced by increasing Sc concentration, the evolution of correlated atomic displacement responses among cations—triggered by the large-amplitude thermal fluctuations of Sc atoms—can play a crucial role in polarization reversal.

## Results

To perform MD simulations under an applied electric field, interatomic forces predicted by the fine-tuned MACE model were combined with electric-field-induced forces derived from BECs predicted by the fine-tuned BM1 model.

The external electric-field-induced force acting on atom $i$, denoted as $F_{\text{ext}}$, was calculated at each MD step following the method proposed by Chen et al.[24,25] according to **Equation (1):**

$$F_{\text{ext}} = |\, e \,|\ \mathcal{E}_{\beta} Z^{*}_{i,\beta\alpha} \tag{1}$$

In Equation (1), $e$ is the elementary charge, $\mathcal{E}_{\beta}$ is the applied electric-field vector, and $Z^{*}_{i,\beta\alpha}$ is the BEC tensor predicted by the equivar_eval model.

The total force acting on each atom during time integration was given as the sum of the interatomic force calculated by MACE and the external electric-field-induced force defined above.

All electric-field-coupled MD simulations were performed under the NPT ensemble. All lattice parameters were allowed to relax under the applied field, in contrast to our previous study in which the in-plane lattice parameters were fixed[28]. The applied electric field was swept cyclically following:

$$0 \rightarrow -20 \rightarrow +20 \rightarrow 0 \text{ MV cm}^{-1}.$$

Prior to the main analysis, we systematically validated both the MLFF models and the simulation setup to ensure quantitative reliability (see Supporting Information S1 and S2 for full details). All validation tests were conducted using a representative composition, $Sc_{0.25}Al_{0.75}N$.

Regarding model accuracy, fine-tuning of the MACE-MP-0 foundation model[26] on the ScAlN-specific PBE dataset dramatically improved the prediction errors to 0.22 meV/atom for energies and 6.4 meV/Å for atomic forces on the held-out test set, outperforming all other state-of-the-art universal MLFFs benchmarked in this study (**Figure S1**, **Table S1**). The BEC prediction model (BM1)[27] similarly achieved mean absolute errors of 0.020 e, 0.011 e, and 0.014 e for Sc, Al, and N atoms, respectively, after fine-tuning, confirming first-principles-level accuracy in the electric-field-induced forces applied during MD (**Figure S3**).

For the simulation setup, we verified that the short-range and long-range models[29] produce negligibly different *polarization-electric field* (*P-E*) hysteresis loops and values of $E_c$ and $P_r$; accordingly, we adopted the computationally efficient short-range model (**Figure S4**(a)). Supercell-size convergence was confirmed by comparing 360-atom and 2880-atom supercells, which yielded $E_c$ values of 9.33 and 8.82 MV cm$^{-1}$, respectively, a difference of less than 6% (Figure S4(b)). Finally, we confirmed that switching from a stepwise to a continuous electric-field ramp protocol does not significantly alter the ferroelectric characteristics, validating the continuous ramp approach adopted for the dynamic analyses presented below (Figure S4(c)). A real-time visualization of the atomic evolution during this continuous electric-field-induced polarization switching is provided in **Movie S1**.

These validations collectively confirm that the present simulation framework is both accurate and well-converged, providing a reliable basis for the quantitative analyses that follow. Having established this foundation, we now examine the dynamic switching behavior of ScAlN, beginning with the sweep-rate dependence of $E_c$. As detailed in Section S2.4, evaluating the dynamic response under varying sweep rates revealed that $E_c$ converges near the quasi-static limit at 0.05 kV cm$^{-1}$ fs$^{-1}$ (**Figure S5**). Because this rate optimally balances physical fidelity with computational cost, it was adopted for all subsequent simulations in this study.

## Ferroelectric Switching Behavior and Composition Dependence

We now turn to the intrinsic ferroelectric switching behavior of $Sc_xAl_{1-x}N$ and its composition dependence. All simulations presented hereafter were performed using a 360-atom supercell with the Short-Range model under a continuous electric-field ramp of 0.05 kV cm$^{-1}$ fs$^{-1}$.

The monotonic reduction of $E_c$ with increasing Sc concentration is among the most reproducible and technologically significant features of ScAlN ferroelectrics, consistently reported across independent experimental studies.[11,30,31] While structural softening and extrinsic effects such as nitrogen vacancies and Sc clustering have been proposed as contributing factors, disentangling their respective roles experimentally is extremely challenging, as film deposition inevitably introduces defects and compositional inhomogeneities that obscure intrinsic switching behavior. Atomistic simulations free from such extrinsic perturbations thus provide a unique opportunity to isolate the intrinsic mechanism. We first examine how the macroscopic *P-E* characteristics evolve with Sc concentration, and then identify the microscopic atomistic origin of the observed $E_c$ reduction.

**Figure 1**(a) shows the calculated *P-E* hysteresis loops for each Sc composition. As the Sc concentration increases,

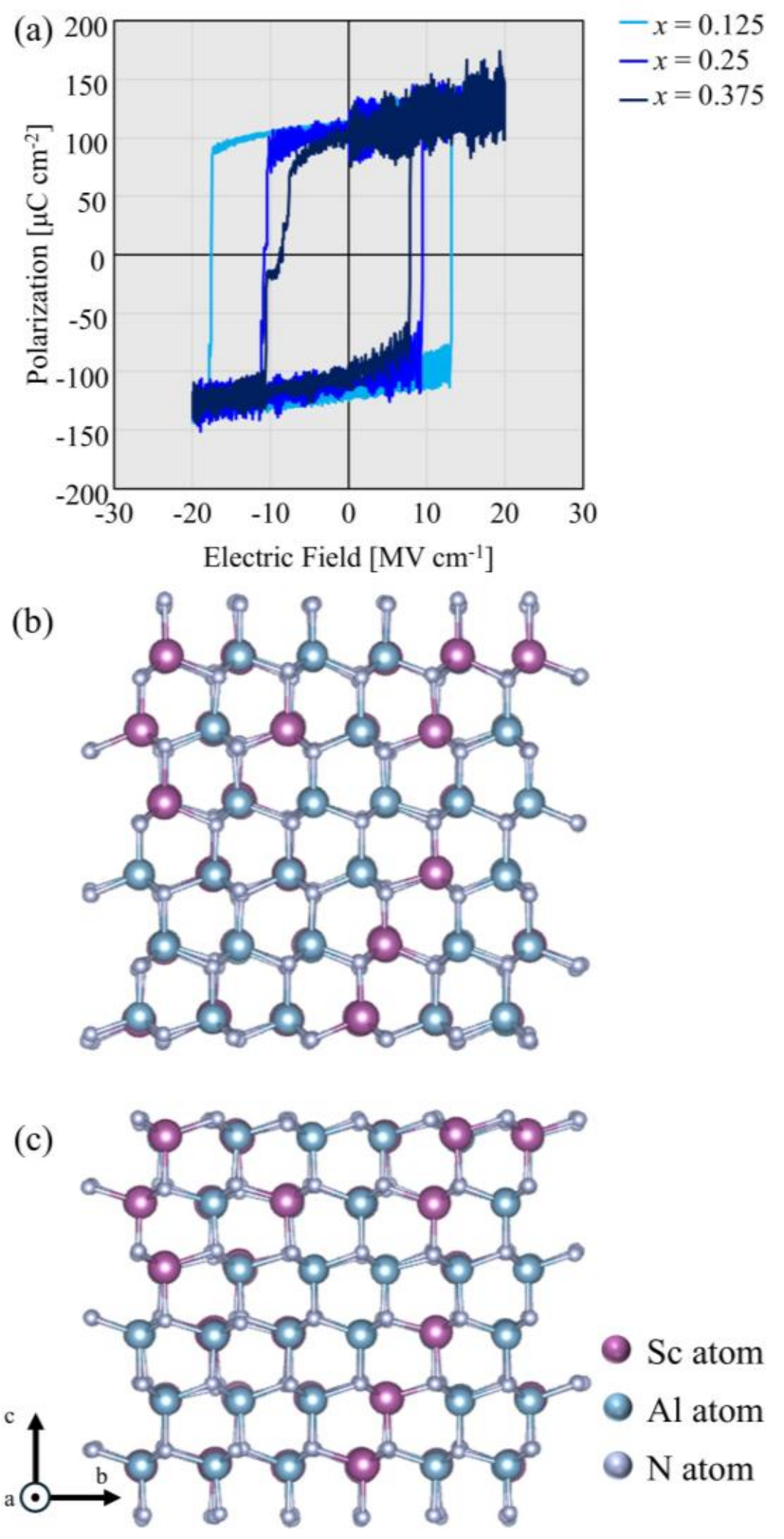

**Figure 1. Composition-dependent ferroelectric switching in ScAlN.** (a) *Polarization-electric field* (*P-E*) hysteresis loops for Sc concentrations of $x = 0.125$ (light blue), 0.25 (blue), and 0.375 (navy). (b) M-polar structure at the positive maximum electric field and (c) N-polar structure at the negative maximum electric field. Structures in panels (b) and (c) are taken from simulations at $x = 0.25$. Purple, blue, and light-violet spheres represent Sc, Al, and N atoms, respectively.

systematic changes are observed in both the polarization switching characteristics and the overall hysteresis shape. To confirm the microscopic origin of the polarization reversal, atomic configurations at the maximum positive and negative electric fields were analyzed. As shown in Figure 1(b) and 1(c), the structures at the maximum positive field correspond to the M-polar configuration, while those at the maximum negative field correspond to the N-polar configuration. Inspection of the atomic displacements clearly confirms the formation of the two opposite polar states in the wurtzite lattice, demonstrating that the present framework correctly reproduces polarization switching between M-polar and N-polar structures.

The present calculations reproduce the monotonic reduction of $E_c$ with increasing Sc concentration over the entire composition range, in good qualitative agreement with multiple experimental reports (Ref. [11,30–34]). In particular, the experimentally observed high-field reduction trend of $E_c$ is consistently captured by the simulations.

The quantitative composition dependence of $E_c$ is further examined in **Figure 2**. Experimentally reported $E_c$ values exhibit substantial scatter across studies, reflecting the strong sensitivity of ferroelectric switching to extrinsic factors, particularly nitrogen vacancy concentration, deposition atmosphere, and local Sc clustering. Films grown under a pure $N_2$ atmosphere, which suppresses vacancy formation and most closely approximates intrinsic switching conditions, yield systematically higher $E_c$ than those deposited under Ar+$N_2$ atmospheres (Ref. [11,31–34]).

The calculated $E_c$ values systematically exceed most of the experimental values, which is physically consistent with the defect-free nature of the SQS simulation model. Among the datasets in Figure 2(a), the pure $N_2$-deposited films of Ref.[30], representing the most intrinsic experimental conditions, show the closest agreement, supporting the interpretation that the calculations capture the upper bound of $E_c$.

The monotonic reduction trend with increasing Sc concentration is robustly reproduced across the entire composition range.

The calculated $P_r$ shows a weaker composition dependence than that reported experimentally, which is consistent with the slight overestimation of the c/a ratio seen in **Figure S2**(a). The slight overestimation of the c/a ratio, a known tendency of PBE-based calculations for nitride systems, leads to a moderated reduction in spontaneous polarization with increasing Sc content.

Nevertheless, the primary objective of this study is to elucidate the atomistic origin of $E_c$ reduction, for which correct reproduction of the monotonic decreasing trend is the essential criterion. The present framework satisfies this criterion and provides a physically consistent basis for the mechanistic analysis that follows.

## Atomistic origins: Sc-triggered switching and evolution of interatomic correlations

Having established the quantitative reliability of the simulation framework and confirmed the correct reproduction of the composition-dependent $E_c$ reduction, we now turn to its atomic-level origin. The central question is why increasing Sc concentration so systematically lowers the switching barrier, a question that static energy analyses alone cannot fully resolve. In the following sections, we address this from two complementary perspectives. We first examine the static structural distortion induced by Sc incorporation, and then analyze the dynamic evolution of interatomic displacement correlations directly extracted from electric-field-induced MD simulations.

## Structural softening driven by internal structural parameters

Figure 2(b) shows the internal structural parameters of Al and Sc for $Sc_xAl_{1-x}N$ at compositions of $x = 0.125$, 0.25, and 0.375. The structural distortion of the wurtzite lattice is characterized by the internal structural parameter $u$, which describes the relative displacement between the cation and anion sublattices along the $c$-axis. In wurtzite-type ferroelectrics, this parameter directly reflects the atomic configuration relevant to polarization switching, because polarization reversal occurs primarily through atomic displacements along the $c$-axis. The parameter is defined as shown in **Equation (2):**

$$u = \frac{z_N - z_{cation}}{c} \quad (2)$$

In Equation (2), $z_N$ and $z_{cation}$ denote the fractional coordinates of the nitrogen atom and the nearest cation (Al or Sc), respectively, and $c$ is the composition-dependent lattice constant along the polarization direction obtained for each specific Sc concentration.

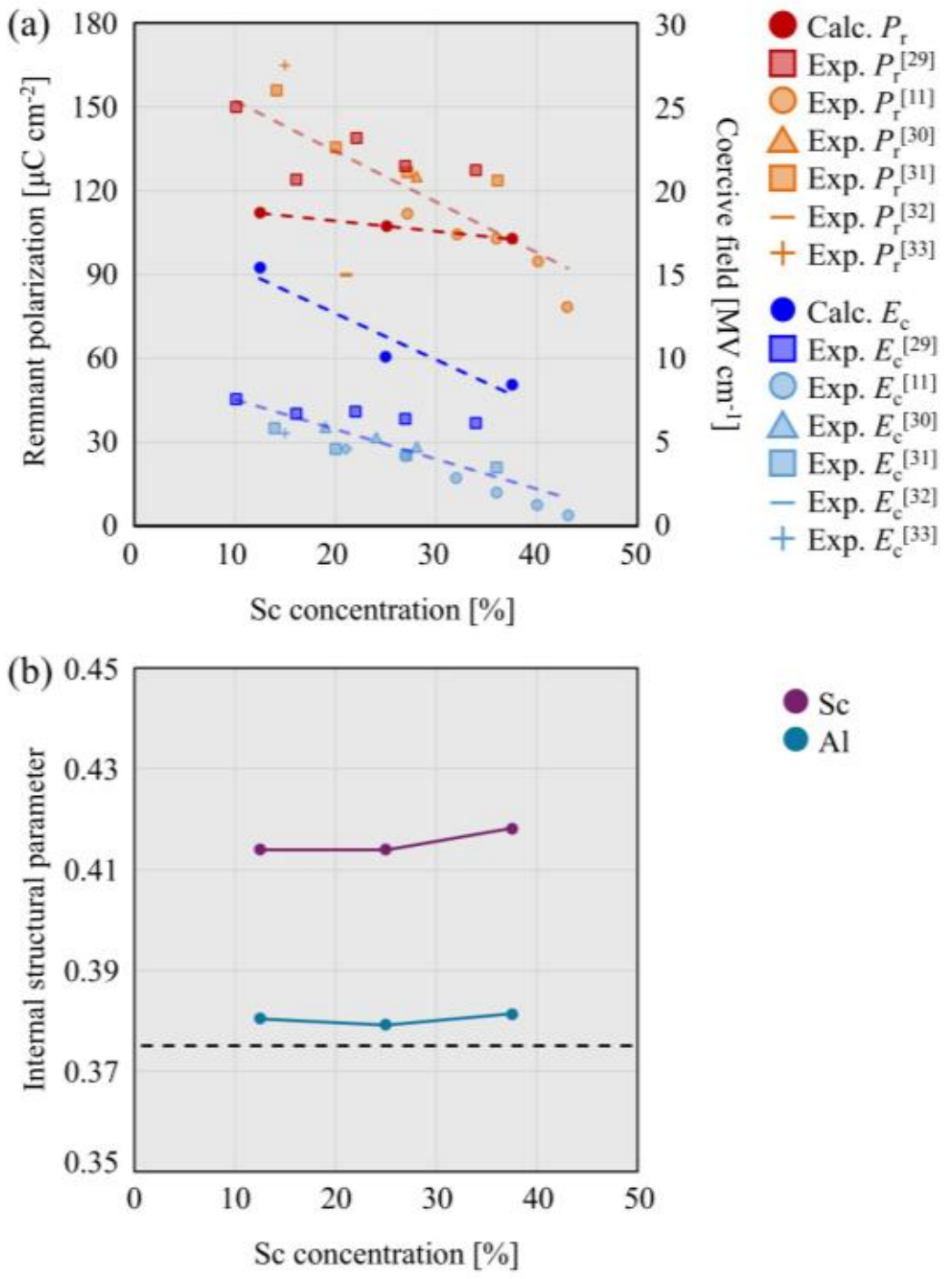

**Figure 2. Composition dependence of ferroelectric properties and the internal structural parameter in ScAlN.** (a) Composition dependence of the remanent polarization ($P_r$) (left axis) and coercive field ($E_c$) (right axis) as a function of Sc concentration. Circles represent the calculated values obtained in this study, while squares indicate experimental data reported in Ref. [11,30–34]. (b) Composition dependence of the internal structural parameter $u$. Values for Sc atoms are shown in purple and those for Al atoms in light blue. The black dashed line indicates the ideal internal structural parameter value of $u$ = 3/8 (0.375) for the wurtzite structure.

Figure 2(b) summarizes the composition dependence of $u$ for Al and Sc atoms. For all compositions studied, Al atoms exhibit values close to the ideal wurtzite parameter $u$ = 3/8 (0.375), whereas Sc atoms systematically show larger deviations from the ideal value. Within the wurtzite switching coordinate, the planar hexagonal nonpolar configuration corresponds to $u = 0.5$. Thus, the larger $u$ values around Sc indicate that the local Sc-N environment is structurally closer to this switching midpoint, whereas the Al-N environment remains closer to the ideal wurtzite configuration. This indicates that the local bonding environment around Sc atoms is inherently more distorted and closer to the configuration associated with polarization reversal.

From a macroscopic perspective, increasing the Sc concentration increases the fraction of locally distorted environments characterized by larger $u$ values. Such deviations from ideal tetrahedral coordination reduce structural anisotropy and weaken the restoring forces opposing polarization reversal. This composition-induced lattice softening is consistent with the experimentally observed decrease in the $c/a$ ratio and has been previously interpreted as a reduction of the static switching barrier on the potential energy surface.

Therefore, static structural softening clearly contributes to the reduction of $E_c$ with increasing Sc concentration. However, the magnitude of the $E_c$ decrease and its systematic nature raise the question of whether static lattice distortion alone is sufficient to account for the observed trend. Because polarization switching occurs under finite temperature and electric field, dynamic atomic motion and cooperative displacement responses may play an additional role beyond the static structural picture.

To address this possibility, we next examine the atomic-scale switching dynamics directly extracted from electric-field-induced MD simulations.

## Dynamic Triggering by Sc Atoms

While static structural softening provides an important contribution, ferroelectric switching is fundamentally a dynamic process under finite temperature and applied electric field. A key question is whether all atomic species respond simultaneously to the applied field, or whether specific atoms initiate the switching event, and if so, whether this sequential response directly lowers the macroscopic switching barrier. To address this, we analyze the time- and field-dependent atomic displacements obtained from electric-field-induced MD simulations. **Figure 3** shows the relationship between the applied electric field and the atomic displacements along the polarization direction ($\Delta z$) for Sc, Al, and N atoms at Sc concentrations of $x = 0.125$, 0.25, and 0.375. For each composition, the Sc, Al, and N displacements are evaluated relative to the corresponding composition-specific planar hexagonal reference defined in the Methods and illustrated in Figure S2(b). Each species exhibits a distinct hysteresis-like displacement-field response. These curves are not independent ferroelectric hysteresis loops, but species-resolved projections of the same collective MD trajectory under an identical electric-field history.

To quantitatively compare the relative progression of these sublattice coordinates during the collective switching event, we define a species-resolved characteristic switching field, $E_{c,atom}$. For each species, the approximately linear stable regions before and after switching were fitted separately, and a midpoint line halfway between the two fitted branches was constructed. $E_{c,atom}$ was then determined from the intersection between the switching portion of the species-resolved displacement curve and this midpoint line. Accordingly, $E_{c,atom}$ is an operational measure of the midpoint of the species-projected displacement response and is not interpreted as an independent coercive field for each atomic species.

The resulting $E_{c,atom}$ values are summarized in **Table 1**.

**Table 1. Species-resolved characteristic switching fields in ScAlN.** The characteristic switching field $E_{c,atom}$ is reported for each atomic species, with values derived from the atomic displacement-electric field data shown in Figure 3.

| Sc concentration | $E_{c,Sc}$ / MV cm$^{-1}$ | $E_{c,Al}$ / MV cm$^{-1}$ | $E_{c,N}$ / MV cm$^{-1}$ |
|---|---|---|---|
| $x = 0.125$ | 13.07 | 13.19 | 13.18 |
| $x = 0.25$ | 10.09 | 10.36 | 10.36 |
| $x = 0.375$ | 7.72 | 7.84 | 7.82 |

Across all compositions examined, Sc atoms consistently exhibit slightly lower $E_{c,atom}$ than Al atoms. For example, at $x = 0.25$, $E_{c,Sc} = 10.09$ MV cm$^{-1}$, whereas $E_{c,Al} = 10.36$ MV cm$^{-1}$. Similar differences are observed at other compositions. This result shows that the Sc displacement response reaches its switching midpoint at a slightly lower field than the Al response within the same collective switching event. Their differences are small in absolute magnitude (0.12-0.27 MV $cm^{-1}$) but are reproduced consistently across all three compositions and other SQS configurations, suggesting a systematic rather than statistical origin. While an absolute difference of 0.12-0.27 MV $cm^{-1}$ might appear modest compared to the total macroscopic $E_c$, its physical significance becomes clear when compared to the intrinsic polarization switching width. By defining the switching onset and completion as 20% and 80% of the normalized polarization transition, the actual switching field widths for $x = 0.125$, 0.25, and 0.375 are found to be extremely narrow (0.275, 0.50, and 0.65 MV $cm^{-1}$, respectively). Thus, the 0.12-0.27 MV $cm^{-1}$ difference accounts for a substantial fraction of the entire switching event, highlighting that Sc initiates displacement significantly earlier than Al within the narrow reversal window. This trend is consistent with Sc atoms initiating displacement reversal marginally earlier than Al atoms under the applied field.

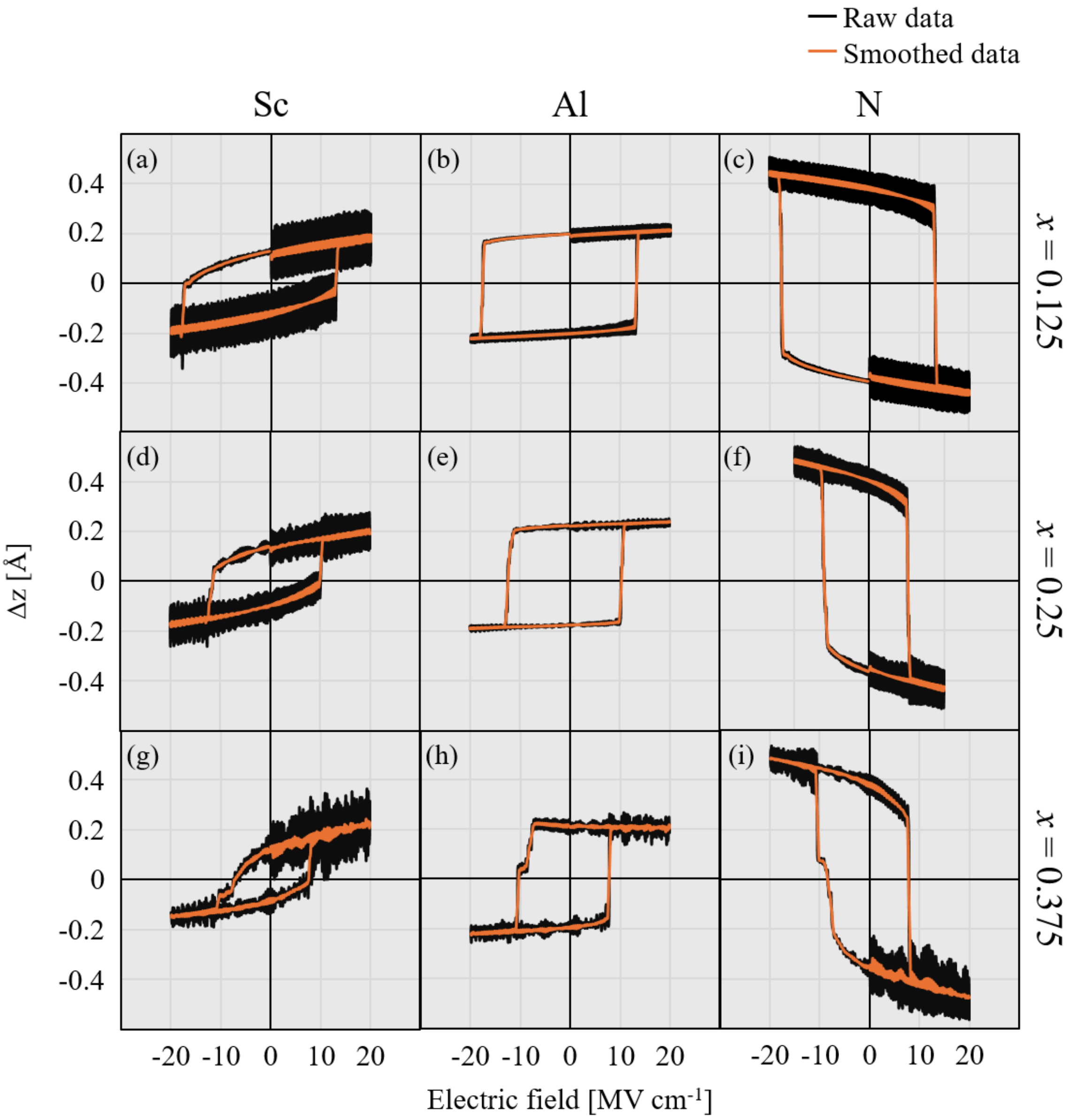

**Figure 3. Relationship between applied electric field and z-direction atomic displacement Δ*z* obtained from electric-field-induced molecular dynamics (MD) simulations.** The Sc, Al, and N curves are species-resolved projections extracted from the same collective MD trajectory rather than independent polarization hysteresis loops. Plots are arranged in a 3×3 matrix, with rows corresponding to Sc concentrations of $x = 0.125$, 0.25, and 0.375 from top to bottom. Columns represent Sc (left), Al (center), and N (right) atoms. Panels (a), (b), and (c) show the Sc, Al, and N responses, respectively, at $x = 0.125$; panels (d), (e), and (f) show the corresponding responses at $x = 0.25$; and panels (g), (h), and (i) show those at $x = 0.375$. Black curves represent the raw atomic displacement data obtained directly from the MD simulations, while orange curves indicate the data after smoothing using a moving average over five consecutive points.

In addition to the earlier onset of displacement, Sc atoms exhibit larger thermal vibration amplitudes compared to Al atoms, as evidenced by the broader spread of $\Delta z$ in Figure 3. This vibrational hierarchy (Al < Sc) indicates that Sc atoms experience a shallower local potential and are more susceptible to field-induced displacement. Together, the earlier displacement onset and enhanced vibrational amplitude are consistent with Sc atoms acting as dynamic triggers for polarization reversal, locally perturbing the lattice and facilitating collective atomic rearrangement.

Importantly, this triggering behavior cannot be inferred solely from static structural parameters. While static distortion lowers the intrinsic energy barrier, the dynamic response reveals how individual atomic species contribute differently to the switching pathway. The earlier displacement of Sc atoms has a direct consequence for the macroscopic switching barrier. Because Sc atoms initiate reversal at a lower field, the remaining Al sublattice is already displaced into a partially switched configuration by the time the bulk switching event occurs. This effectively reduces the energy cost for the subsequent collective Al displacement, lowering the apparent *Ec*. The enhanced thermal amplitude of Sc atoms further amplifies this effect: larger vibrational excursions increase the probability that thermal fluctuations alone carry Sc atoms across the local potential barrier, allowing field-assisted switching to commence at a smaller applied field.

However, polarization reversal in a crystalline solid is inherently cooperative. The displacement of one atomic species must be accommodated by correlated motion of neighboring atoms. Therefore, to fully understand how Sc incorporation reduces $E_c$, it is necessary to examine how these local dynamic perturbations propagate through the lattice via interatomic displacement correlations. This issue is addressed in the following section.

**Evolution of Interatomic Displacement Correlations and Unified Mechanism**

To understand how the local dynamic instability of Sc atoms propagates through the lattice and contributes to macroscopic switching behavior, we performed a correlation analysis of atomic displacements during electric-field-induced MD simulations.

Specifically, for each composition, we evaluated the linear correlation between the z-direction displacement of Sc atoms ($\Delta z_{Sc}$) and those of neighboring Al and N atoms per electric-field step. The resulting correlation slopes quantify both the strength and phase relationship of cooperative atomic response under the applied electric field.

**Figure 4** shows the displacement correlations for different Sc concentrations. For all compositions, N atoms exhibit a negative correlation with Sc displacements, indicating a robust anti-phase response consistent with the alternating cation-anion displacement characteristic of polarization reversal. The slopes for N remain relatively stable across compositions, suggesting that the anion framework plays a structurally constraining role during switching.

In contrast, the behavior of Al atoms evolves systematically with increasing Sc concentration. At low Sc concentration ($x = 0.125$), Al displacements are positively correlated with Sc displacements, indicating in-phase cooperative motion. As the Sc content increases, the magnitude of this positive slope decreases. At higher Sc concentration ($x = 0.375$), the slope becomes negative, revealing a qualitative

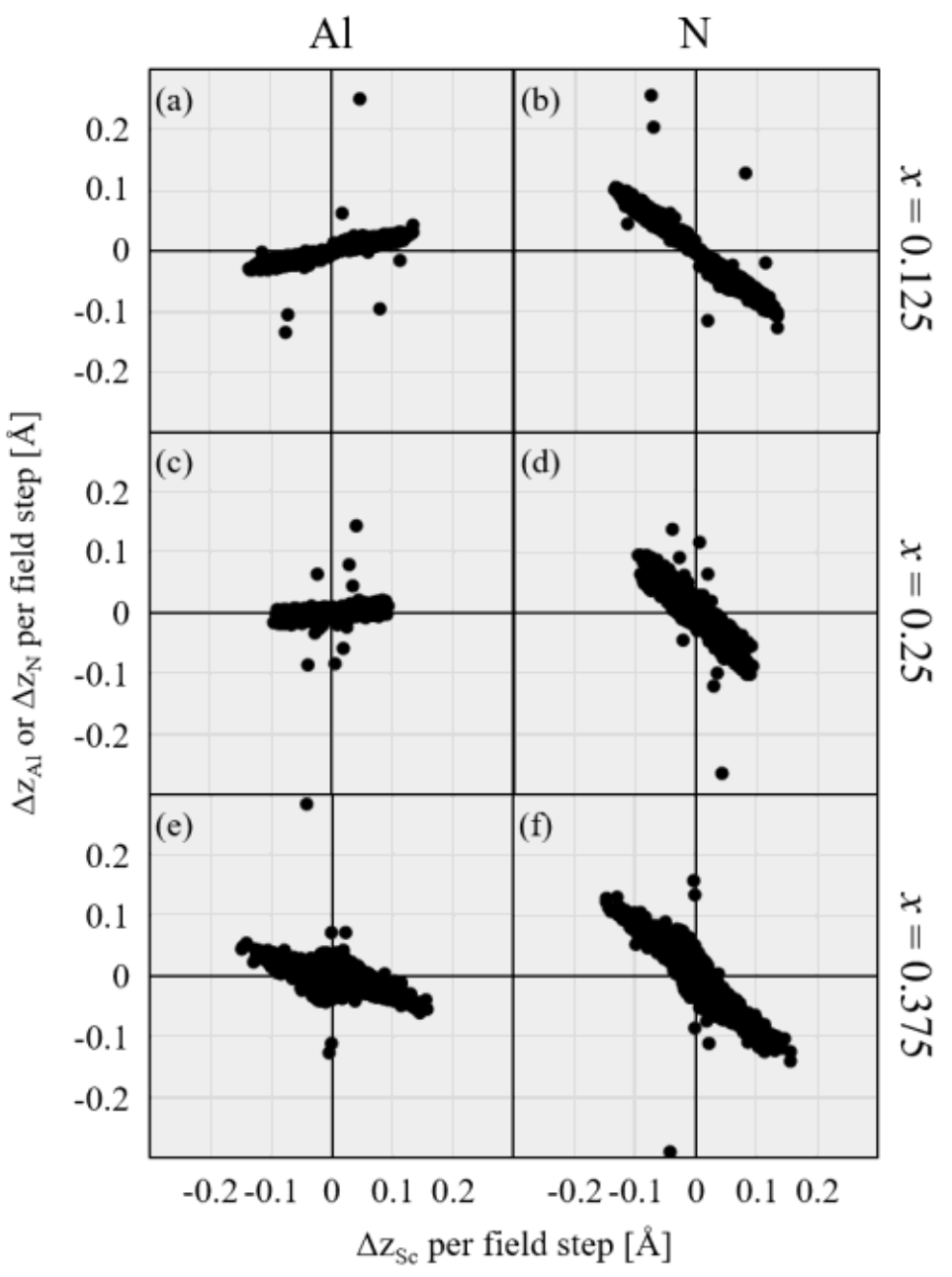

**Figure 4. Composition-dependent correlation analysis of atomic displacements during polarization switching.** Correlations between Sc atomic displacement per electric-field step and $z$-direction displacements of Al atoms (left column) and N atoms (right column). Plots are arranged in a 3×2 matrix, with rows corresponding to Sc concentrations of $x$ = 0.125, 0.25, and 0.375 from top to bottom. Panels (a) and (b) show the Sc-Al and Sc-N correlations, respectively, at $x$ = 0.125; panels (c) and (d) show the corresponding correlations at $x$ = 0.25; and panels (e) and (f) show those at $x$ = 0.375. Black points represent the displacement pairs sampled at each electric-field step.

transition to partially out-of-phase motion between Al and Sc atoms.

This sign reversal of the Al-Sc displacement correlation represents a fundamental change in cooperative response mode, and its physical origin can be traced to the difference in bond strength between Sc-N and Al-N pairs. Crystal orbital Hamilton population (COHP) analysis yields average -ICOHP values of around 3.0 eV for Sc-N bonds and 4.7 eV for Al-N bonds, demonstrating that Sc-N bonds are weaker. This bond-strength hierarchy results in different local mechanical compliances for the two cation species: Sc atoms occupy a shallower potential well, which accounts for both their larger thermal vibration amplitudes and their earlier onset of field-induced displacement established in the "Dynamic Triggering by Sc Atoms" subsection of the Results.

Although both Sc and Al possess positive BECs and are directly driven along the same applied-field direction, their relative displacements are not determined by the sign of this driving force alone. Rather, they form part of a mechanically coupled Sc-Al-N network, and their relative displacement response is governed by how this network collectively accommodates the applied field. As Sc concentration increases, the fraction of weaker Sc-N bonds in the lattice grows, progressively altering the mechanical coupling between cation sublattices. At low Sc concentration, the rigid Al-N network dominates, constraining Sc displacements into synchronized, in-phase motion with Al. At higher Sc concentration, the increasing prevalence of compliant Sc-N bonds weakens this constraint: Sc atoms can undergo larger displacements relative to Al, allowing Sc to respond more independently of the Al-N framework.

This composition-dependent change in local mechanical coupling reorganizes the collective displacement response and provides a physical basis for the observed evolution from in-phase to partially out-of-phase Al-Sc correlation. Critically, this partially decoupled response enables a sequential switching pathway: Sc atoms, driven by their weaker bonding environment, initiate displacement reversal first, lowering the local energy landscape for the subsequent Al displacement. This sequential process carries a lower cooperative energy cost than the fully concerted switching required at low Sc concentration, directly contributing to the reduction of macroscopic $E_c$.

The evolution of displacement correlation directly parallels the monotonic reduction of $E_c$ observed in Figure 2(a). The quantitative correspondence between the sign change of the Al-Sc correlation slope and the composition-dependent $E_c$ reduction strongly suggests that this reorganization of cooperative displacement modes is a governing factor in the switching

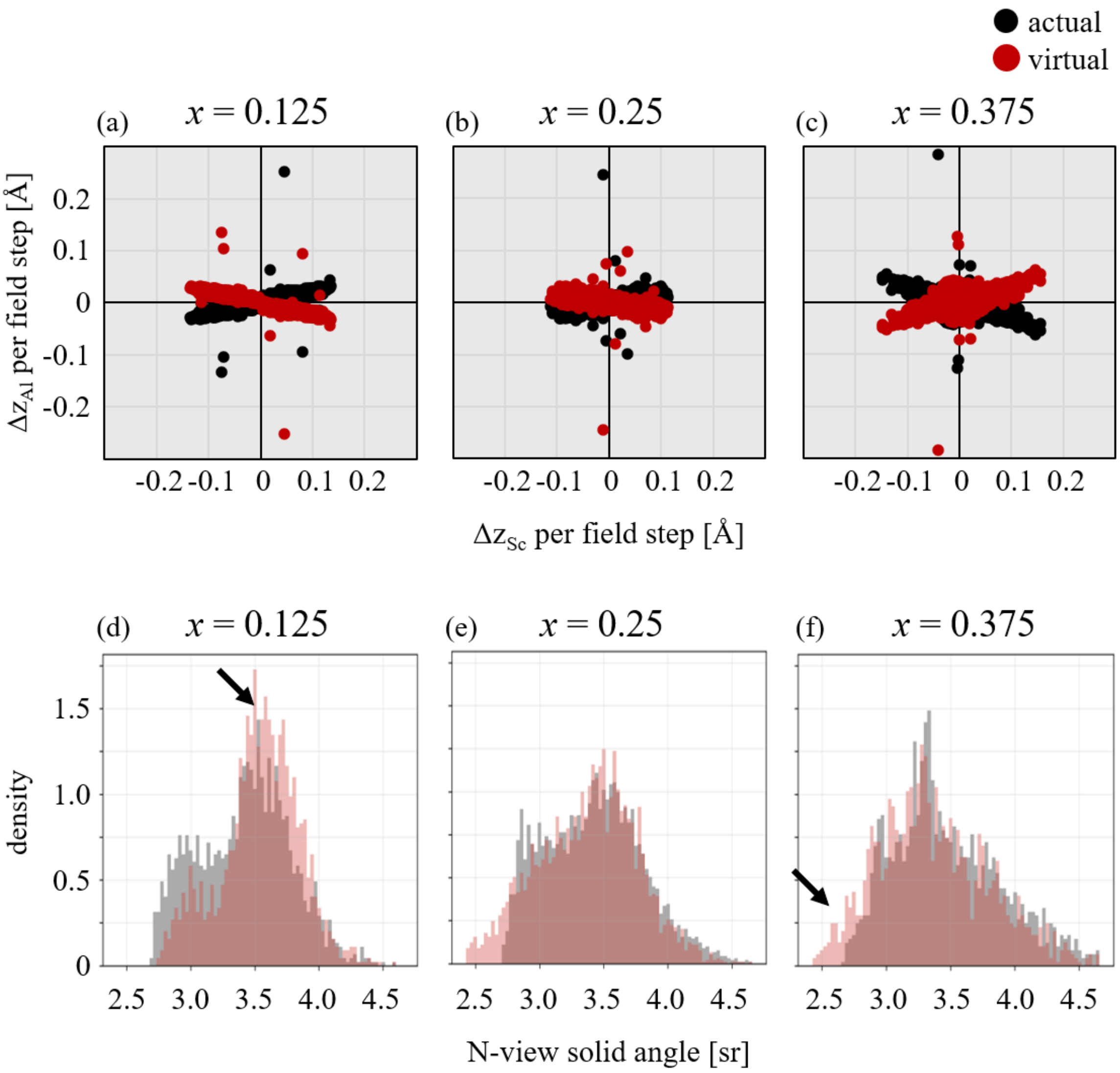

**Figure 5. Effect of the Sc-Al displacement phase relationship on the local cation-gate geometry.** (a-c) Correlations between Sc and Al atomic displacements for the actual molecular dynamics trajectories (black) and the generated virtual trajectories (red) at Sc concentrations of $x = 0.125$, 0.25, and 0.375, respectively. The virtual trajectories are generated by artificially reversing the Al displacement direction at each step. (d-f) Histograms of the solid angle of the cation triangle viewed from the N atom for the actual (black) and virtual (red) structures at $x = 0.125$, 0.25, and 0.375, respectively. The vertical axis represents density. The arrows highlight key differences in the distribution profiles discussed in the text, specifically near 3.5 sr at $x = 0.125$ and 2.5 sr at $x = 0.375$, where sr denotes steradian.

barrier, operating in addition to static structural softening.

To examine the geometrical consequence of this composition-dependent reorganization of the collective cation response, we analyzed the local steric environment experienced by N during switching. Specifically, we focused on the solid angle of the "cation triangle" (formed by the three cations through which the N atom must pass during the switching process) as viewed from the N atom. A larger solid angle indicates a wider passage for the N atom, meaning lower steric hindrance during polarization switching.

To isolate the effect of this dynamic phase relationship on the solid angle, we created a virtual trajectory by artificially reversing only the displacement direction of the Al atoms based on the actual molecular dynamics trajectory. The virtual coordinates of the Al atoms at each step were calculated as $Position'(\mathrm{n}) = Position(\mathrm{n}-1) - \{Position(\mathrm{n}) - Position(\mathrm{n}-1)\}$. By performing this calculation using relative displacements at each step, we can generate a virtual structure with an inverted phase relationship (in-phase vs. out-of-phase) between Sc and Al displacements, while minimizing the accumulation of unnatural strain caused by the artificial movement of Al.

**Figure 5**(a-c) shows the correlation between the actual trajectory (black) and the virtual trajectory (red) for each composition. This operation clearly inverts the phase relationship. Specifically, at $x$ = 0.125, the original "in-phase" relationship (black) is inverted to "out-of-phase" in the virtual structure (red). At $x$ = 0.25, it is similarly inverted, although the magnitude of the correlation slope is intrinsically small. Conversely, at $x$ = 0.375, the original "out-of-phase" relationship (black) is inverted to "in-phase" in the virtual structure (red).

Subsequently, Figure 5(d-f) shows the histograms of the solid angle for each composition. The solid angle analysis reveals a trend that is fully consistent with the composition-dependent Ec reduction shown in Figure 2(a), where the largest single drop in Ec occurs between $x$ = 0.125 and $x$ = 0.25, while the further reduction from $x$ = 0.25 to $x$ = 0.375 is comparatively modest.

At $x$ = 0.125, where the in-phase Al-Sc correlation is strongest and the Ec is largest, the effect of the phase relationship on steric hindrance is most pronounced. The actual in-phase structure (black) exhibits a prominent peak shifted toward the lower solid angle side (~3.0-3.2 sr) with reduced density near 3.5 sr, whereas the virtual out-of-phase structure (red) suppresses this low-angle peak and maintains a systematically larger solid angle.

This demonstrates that the in-phase constraint actively narrows the cation gate, imposing an additional steric cost on polarization switching. The large $E_c$ at $x$ = 0.125 is thus attributable not only to the static structural rigidity of the Al-N network, but also to this dynamic gating penalty imposed by the in-phase correlation.

At $x$ = 0.25 and 0.375, the solid angle distributions of the actual and virtual structures converge, showing only marginal differences. This convergence is physically consistent with the modest $E_c$ reduction observed between these two compositions in Figure 2(a).

At $x$ = 0.25, the Al-Sc correlation slope is already small, meaning the cooperative constraint has largely dissolved; inverting Al displacements therefore has little geometric consequence. At $x$ = 0.375, while the out-of-phase correlation is established, the Al displacement amplitude is inherently smaller relative to Sc as a consequence of the mechanical decoupling discussed above; thus, Al reversal produces only a limited change in the overall cation-triangle geometry.

In addition to these distribution-based features, we quantified the mean solid angle of the actual MD trajectories to directly assess its composition dependence. The mean solid angles are 3.418, 3.423, and 3.475 sr for $x$ = 0.125, 0.25, and 0.375, respectively, indicating a modest overall increase with increasing Sc concentration. The corresponding mean and standard deviation for both the actual and virtual trajectories are summarized in **Table S2** of the Supporting Information. These averaged values provide a complementary quantitative measure of the overall gate size, whereas the full distributions in Figure 5(d-f) capture the detailed redistribution of local cation-gate geometries associated with the Sc-Al displacement phase relationship.

In both cases, the reduced sensitivity of the solid angle to the phase relationship reflects the fact that the contribution of dynamic correlations has already largely diminished by $x$ = 0.25, and the further $E_c$ reduction toward $x$ = 0.375 is driven predominantly by continued static structural softening.

The solid angle analysis thus provides a geometrically grounded account of the composition-dependent interplay between dynamic and static contributions to $E_c$ reduction. The clearest evidence is at $x$ = 0.125: the in-phase constraint demonstrably narrows the cation gate, and its progressive dissolution with increasing Sc concentration constitutes a physically distinct contribution to $E_c$ reduction that operates in addition to, and independently of, static structural softening. The geometrical picture emerging from this analysis is summarized schematically in **Figure 6**. The evolution from predominantly in-phase toward partially out-of-phase Sc-Al displacement can modify the local cation-gate geometry, increasing the solid angle viewed from N and thereby reducing the steric hindrance encountered by N during polarization reversal. Taken together, the results reveal a unified mechanism for $E_c$ reduction in ScAlN. Static structural softening, characterized by the deviation of the internal structural parameter $u$, lowers the intrinsic polarization-reversal barrier. Simultaneously, dynamic correlation evolution modifies how atomic displacements propagate through the lattice under finite temperature and electric field. The synergy between these static and dynamic effects reduces the effective switching barrier and gives rise to the pronounced composition-dependent decrease in $E_c$.

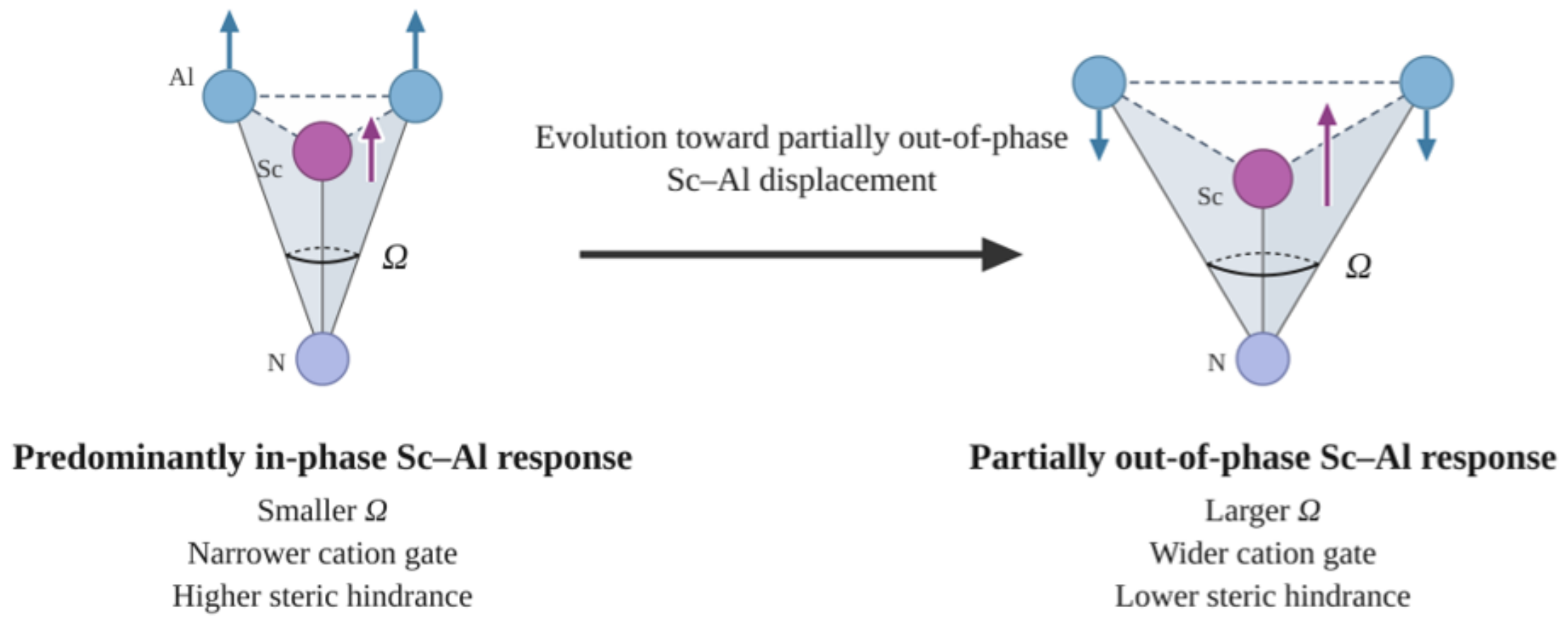

**Figure 6. Schematic illustration of the geometrical consequence of the Sc-Al displacement phase relationship during polarization switching.** A predominantly in-phase Sc-Al response is associated with a smaller solid angle Ω subtended by the cation triangle as viewed from the N atom, corresponding to a narrower cation gate and higher steric hindrance. When Sc and Al are constrained to move in phase (left), the cation triangle moves rigidly, and the gate viewed from N remains narrow. When Sc is released from this constraint and leads the switching, the triangle deforms and the gate widens, reducing the steric hindrance encountered by N during polarization reversal. Purple, blue, and light-violet spheres represent Sc, Al, and N atoms, respectively. Ω denotes the solid angle subtended by the cation triangle as viewed from the N atom. The arrows adjacent to the cations schematically indicate the displacement directions of Sc and Al, while the arrow through the cation triangle indicates the N switching pathway. This is a conceptual illustration; the atomic displacements and geometric opening are schematic and not drawn to scale, and the diagram is not intended to imply that all out-of-phase responses produce an identical change in the cation-gate geometry.

Importantly, this mechanism extends beyond a purely energetic description on a static potential energy surface. Instead, it highlights the essential role of dynamic correlation engineering—the systematic modification of cooperative atomic displacement patterns—as a governing factor in ferroelectric switching.

This dynamic perspective provides a concrete design criterion for hexagonal ferroelectrics that goes beyond the conventional static-barrier picture: an effective dopant must simultaneously satisfy two conditions. First, it must locally soften the host lattice by deviating the internal structural parameter $u$ toward the transition-state configuration (structural softening condition). Second, its bonding to the anion framework must be weak enough to release the in-phase constraint imposed by the host cation network, allowing the dopant to lead the switching within the experimentally accessible composition range (dynamic correlation condition). Candidate dopants can be screened against these two criteria using the simulation framework established here, offering a computationally efficient route to identify substituents that optimize the trade-off between switching field and polarization in hexagonal ferroelectrics.

## Discussion

In this study, we employed a first-principles-accurate simulation framework for wurtzite ferroelectric $Sc_xAl_{1-x}N$ by integrating a fine-tuned MLFF with an equivariant neural-network-based BEC model. This approach enabled large-scale electric-field-induced MD simulations capable of quantitatively reproducing experimentally observed trends in lattice anisotropy and $E_c$ reduction.

We demonstrated that $E_c$ decreases monotonically with increasing Sc concentration and clarified its microscopic origin from both static and dynamic perspectives. Static structural softening, characterized by systematic deviation of the internal structural parameter $u$, lowers the intrinsic polarization-reversal barrier. Beyond this static picture, we revealed that Sc incorporation induces a qualitative evolution of cation-cation displacement correlations during switching. This evolution is rooted in the weaker Sc-N bonding relative to Al-N, as quantified by COHP analysis (-ICOHP: approximately 3.0 and 4.7 eV for Sc-N and Al-N, respectively), which places Sc atoms in a shallower local potential and renders them more susceptible to field-induced displacement. Sc atoms exhibit enhanced thermal instability and initiate displacement reversal earlier than Al atoms, while the correlation between Sc and Al displacements transitions from predominantly in-phase to partially out-of-phase motion with increasing Sc concentration. This mechanical decoupling enables a sequential switching pathway in which Sc atoms initiate reversal first, reducing the cooperative energy cost for the subsequent Al displacement and lowering the effective switching barrier under applied electric field.

These results demonstrate that $E_c$ reduction in ScAlN cannot be understood solely within a static potential-energy framework. Instead, it arises from the synergy between structural softening and dynamic correlation evolution, the latter being driven by the composition-dependent reorganization of mechanical coupling between cation sublattices. The concept highlighted in this work—engineering cooperative atomic response modes to lower switching barriers—provides a dynamic design principle for hexagonal ferroelectrics and offers guidance for developing low-voltage ferroelectric materials.

## Methods

### Construction of the ScAlN Database

To fine-tune a MACE model specialized for the ScAlN system, we employed a first-principles dataset publicly available from Ref. [35].

The reference calculations were performed using the Vienna Ab initio Simulation Package (VASP)[36–39] based on the projector augmented-wave (PAW) method.[40] The dataset consists of 256-atom supercell structures of $Sc_xAl_{1-x}N$ calculated using the PBE exchange-correlation functional,[41] with four Sc compositions: $x = 0$, 0.125, 0.25, and 0.375.

The dataset contains a total of 1664 structures. To rigorously evaluate the generalization performance of the model, the dataset was randomly divided into a training set (1332 structures) and a test set (332 structures). The predictive accuracy on the test set, which was not used during training, was used to assess the reliability of the model.

In the reference calculations, the PBE exchange-correlation functional was employed, and Brillouin-zone integration was restricted to the Gamma point (1×1×1). From this public dataset, the total energies, atomic forces, and stress tensors were extracted and used for training the MLFF. To investigate the bond strength, ICOHP was calculated using Lobster code[42–44].

### Development of the MACE Model

To improve the modeling accuracy for ScAlN, we fine-tuned the general-purpose

MLFF MACE-MP-0[26] using the PBE-based dataset described in Construction of the ScAlN Database subsection of the Methods. MACE-MP-0 served as a foundation model, and the fine-tuning process enabled adaptation to the structural and energetic characteristics specific to ScAlN.

The dataset split followed the procedure described in Construction of the ScAlN Database subsection of the Methods. Training was performed using a batch size of 2 and an initial learning rate of 0.01 with the AMSGrad optimizer. During the early stage of training, an exponential moving average (EMA) of model weights was applied to stabilize convergence. After the validation loss reached a plateau, stochastic weight averaging (SWA) was introduced. Early stopping was applied when the validation loss failed to improve for five consecutive epochs.

The total loss, defined as an equally weighted sum of energy and atomic force errors, decreased consistently throughout the training process. As a result, a model with high predictive accuracy on the test dataset was obtained.

**Born Effective Charge Calculation**

Most existing MLFFs are trained solely on structural and energetic information and do not explicitly incorporate electronic properties. This limitation becomes critical for systems in which electric-field effects play a central role. To quantitatively evaluate polarization changes induced by atomic displacements and to directly account for electric-field-induced forces, the introduction of BECs is essential.[45–47]

In this study, we adopted the equivar_eval model (architecture: BM1), an equivariant graph convolutional neural network developed by Kutana et al.[27]

The same structural dataset used for MACE training (Train: 1332, Test: 332) was employed for training and testing the BEC prediction model. Since the original public dataset does not provide reference BEC values, BECs were independently calculated in this study using density functional perturbation theory (DFPT). DFPT calculations were performed using the VASP code, employing the PBE functional and Gamma-point sampling to ensure consistency with the force-field training data, and were carried out for fully relaxed structures.

Fine-tuning of the BEC model was conducted using a batch size of 16, a learning rate of $5\times10^{-3}$, and the AdamW optimizer. Training continued for up to 300 epochs, with early stopping applied if no improvement was observed for 20 epochs. The loss function was defined as the L1 loss between predicted and DFPT reference BEC values. Through this procedure, a BEC prediction model with substantially reduced test-set error was obtained, providing sufficient accuracy for subsequent electric-field-induced MD simulations.

**Polarization Evaluation and Construction of the *P-E* Hysteresis Loop**

To evaluate the electric polarization under an external electric field and to construct *P-E* hysteresis loops, the polarization was calculated at each electric-field step during the electric-field-induced MD simulations. The polarization $P$ was evaluated using atomic displacements and BECs according to **Equation (3)**:

$$\boldsymbol{P} = \frac{\boldsymbol{e}}{\boldsymbol{V}} \sum_{i} \boldsymbol{Z}_i^* \, \boldsymbol{\Delta z}_i \qquad (3)$$

In Equation (3), $e$ is the elementary charge, $V$ is the instantaneous cell volume, $\boldsymbol{Z}_i^*$ denotes the BEC of atom $i$ predicted by the fine-tuned ML model, and $\boldsymbol{\Delta z}_i$ represents the displacement of atom $i$ along the polarization direction relative to the composition-specific planar hexagonal nonpolar reference structure.

For each Sc composition, a composition-specific planar hexagonal reference was constructed from the corresponding SQS supercell. The Sc/Al occupations, in-plane atomic coordinates, and lattice vectors were retained, while each N atom was displaced along the $\boldsymbol{c}$ direction into the nearest cation plane. This construction represents the planar hexagonal midpoint of the wurtzite switching coordinate and is used as the zero of the relative polarization and atomic-displacement analyses. Because the random Sc/Al occupation in the finite alloy supercell does not necessarily preserve exact inversion symmetry, we refer to this configuration as a planar hexagonal nonpolar reference rather than a strictly centrosymmetric structure. The reference configuration is illustrated in Figure S2(b).

Atomic displacements were evaluated relative to this reference using the minimum-image convention and projected onto the instantaneous ***c*** axis. The same reference definition was used consistently for the polarization evaluation and for the species-resolved displacement analysis shown in Figure 3. Atomic displacements relative to this reference structure were combined with the corresponding BEC values to compute the polarization using Equation (3).

At each electric-field step, the system was equilibrated for 20 ps, followed by a 10 ps sampling period during which atomic positions were recorded. Time-averaged atomic displacements were calculated from this trajectory, and the average polarization was obtained accordingly. The *P*-*E* hysteresis loop was constructed by plotting the averaged polarization as a function of the applied electric field.

This procedure ensures a consistent evaluation of polarization based on dynamically computed atomic displacements and fine-tuned BECs, enabling accurate and efficient modeling of ferroelectric switching behavior within the MLFF-based MD framework.

To assess the robustness of the simulation setup, the influence of the interaction range was evaluated by comparing the default Short Range MACE model with a locally extended scheme (LES)-based Long Range model[29] using a 2880-atom supercell. Finite-size effects were evaluated by comparing 360-atom and 2880-atom supercells. The influence of the electric-field application scheme was evaluated by comparing a quasi-static stepwise protocol with field increments of 1 MV $cm^{-1}$ and the continuous-ramp protocol used for the switching-dynamics analyses. In addition, the sweep-rate dependence was examined at 0.25, 0.05, and 0.01 kV $cm^{-1}$ $fs^{-1}$ for $Sc_{0.25}Al_{0.75}N$. The corresponding validation results are presented in Figures S4 and S5 and Movie S1 of the Supporting Information.

To properly model the disordered solid solution state, the Sc-doped supercells (360 atoms) used in these MD simulations were constructed based on the SQS method[48]. Since previous studies have indicated that the distribution of Sc in ScAlN is largely random, the SQS approach provides a physically appropriate representation.[13,49,50] To quantitatively evaluate the randomness of the SQS configurations on the cation sublattice, a spin-like variable $\sigma_i = +1$ for Sc and $\sigma_i = -1$ for Al was assigned to each cation site. The pair correlation function for the *f*-th nearest-neighbor shell was evaluated as $\Pi_f = \langle\sigma_i\ \sigma_j\rangle_f$. For a perfectly random solid solution, the expected value is $\Pi_f^{random} = (2x-1)^2$, where $x$ is the Sc concentration. The deviation from the ideal random distribution was quantified as $\Delta_f = |\Pi_f - \Pi_f^{random}|$. Furthermore, to verify the effect of local atomic configuration on the macroscopic ferroelectric properties, we performed identical field-induced MD simulations on alternative SQS configurations for each Sc composition. As detailed in the Supporting Information (**Figure S6**), the configuration dependency was found to be minimal; the variations in $E_c$ and $P_r$ between different atomic arrangements are substantially smaller than the composition-dependent trends. This confirms that the pronounced composition dependence observed in this study is highly robust against local structural variations.

**Acknowledgements**

The computation was carried out using the computer resources offered by the Research Institute for Information Technology, Kyushu University.

**Funding**

This study was supported by the Ministry of Education, Culture, Sports, Science and Technology (MEXT, 26K0120) and the New Energy and Industrial Technology Development Organization (NEDO). P.-Y.C. acknowledges support from JST SPRING (Grant Number JPMJSP2108). The funders had no role in the study design, data collection and analysis, interpretation of the results, or preparation of the manuscript.

**Competing interests**

The authors declare no competing financial or non-financial interests.

**Author Contributions**

R.S. and P.-Y.C. conceived the idea and designed the research. R.S. performed the simulations, analyzed the data, and prepared the figures. R.S. wrote the original draft of the manuscript. T.M. supervised the project and acquired funding. All authors discussed the results and reviewed and edited the manuscript.

**Data Availability**

The fine-tuned Born effective charge model (BM1.pt), a test structural dataset in xyz format, and the trained MACE force-field model used in this study are freely available in the GitHub repository at https://github.com/nmdl-mizo/ScAlN_BEC_verification. These publicly available files are provided to facilitate the reproduction and verification of the results reported in this study.

**Code availability**

The inference script used in this study is freely available in the GitHub repository at https://github.com/nmdl-mizo/ScAlN_BEC_verification.

**Figure legends**

**Figure 1. Composition-dependent ferroelectric switching in ScAlN.** (a) *Polarization-electric field* (*P-E*) hysteresis loops for Sc concentrations of $x$ = 0.125 (light blue), 0.25 (blue), and 0.375 (navy). (b) M-polar structure at the positive maximum electric field and (c) N-polar structure at the negative maximum electric field. Structures in panels (b) and (c) are taken from simulations at $x$ = 0.25. Purple, blue, and light-violet spheres represent Sc, Al, and N atoms, respectively.

**Figure 2. Composition dependence of ferroelectric properties and the internal structural parameter in ScAlN.** (a) Composition dependence of the remanent polarization ($P_r$) (left axis) and coercive field ($E_c$) (right axis) as a function of Sc concentration. Circles represent the calculated values obtained in this study, while squares indicate experimental data reported in Ref. [11,30–34]. (b) Composition dependence of the internal structural parameter $u$. Values for Sc atoms are shown in purple and those for Al atoms in light blue. The black dashed line indicates the ideal internal structural parameter value of $u$ = 3/8 (0.375) for the wurtzite structure.

**Figure 3. Relationship between applied electric field and z-direction atomic displacement $\Delta z$ obtained from electric-field-induced molecular dynamics (MD) simulations.** The Sc, Al, and N curves are species-resolved projections extracted from the same collective MD trajectory rather than independent polarization hysteresis loops. Plots are arranged in a 3×3 matrix, with rows corresponding to Sc concentrations of $x$ = 0.125, 0.25, and 0.375 from top to bottom. Columns represent Sc (left), Al (center), and N (right) atoms. Panels (a), (b), and (c) show the Sc, Al, and N responses, respectively, at $x$ = 0.125; panels (d), (e), and (f) show the corresponding responses at $x$ = 0.25; and panels (g), (h), and (i) show those at $x$ = 0.375. Black curves represent the raw atomic displacement data obtained directly from the MD simulations, while orange curves indicate the data after smoothing using a moving average over five consecutive points.

**Figure 4. Composition-dependent correlation analysis of atomic displacements during polarization switching.** Correlations between Sc atomic displacement per electric-field step and $z$-direction displacements of Al atoms (left column) and N atoms (right column). Plots are arranged in a 3×2 matrix, with rows corresponding to Sc concentrations of $x$ = 0.125, 0.25, and 0.375 from top to bottom. Panels (a) and (b) show the Sc-Al and Sc-N correlations, respectively, at $x$ = 0.125; panels (c) and (d) show the corresponding correlations at $x$ = 0.25; and panels (e) and (f) show those at $x$ = 0.375. Black points represent the displacement pairs sampled at each electric-field step.

**Figure 5. Effect of the Sc-Al displacement phase relationship on the local cation-gate geometry.** (a-c) Correlations between Sc and Al atomic displacements for the actual molecular dynamics trajectories (black) and the generated virtual trajectories (red) at Sc concentrations of $x = 0.125$, 0.25, and 0.375, respectively. The virtual trajectories are generated by artificially reversing the Al displacement direction at each step. (d-f) Histograms of the solid angle of the cation triangle viewed from the N atom for the actual (black) and virtual (red) structures at $x = 0.125$, 0.25, and 0.375, respectively. The vertical axis represents density. The arrows highlight key differences in the distribution profiles discussed in the text, specifically near 3.5 sr at $x = 0.125$ and 2.5 sr at $x = 0.375$, where sr denotes steradian.

**Figure 6. Schematic illustration of the geometrical consequence of the Sc-Al displacement phase relationship during polarization switching.** A predominantly in-phase Sc-Al response is associated with a smaller solid angle Ω subtended by the cation triangle as viewed from the N atom, corresponding to a narrower cation gate and higher steric hindrance. When Sc and Al are constrained to move in phase (left), the cation triangle moves rigidly, and the gate viewed from N remains narrow. When Sc is released from this constraint and leads the switching, the triangle deforms and the gate widens, reducing the steric hindrance encountered by N during polarization reversal. Purple, blue, and light-violet spheres represent Sc, Al, and N atoms, respectively. Ω denotes the solid angle subtended by the cation triangle as viewed from the N atom. The arrows adjacent to the cations schematically indicate the displacement directions of Sc and Al, while the arrow through the cation triangle indicates the N switching pathway. This is a conceptual illustration; the atomic displacements and geometric opening are schematic and not drawn to scale, and the diagram is not intended to imply that all out-of-phase responses produce an identical change in the cation-gate geometry.

**Table 1. Species-resolved characteristic switching fields in ScAlN.** The characteristic switching field $E_{c,atom}$ is reported for each atomic species, with values derived from the atomic displacement-electric field data shown in Figure 3.

# Supporting Information

## Supporting Information for: Origin of Reduced Coercive Field in ScAlN: Synergy of Structural Softening and Dynamic Atomic Correlations

[1]Ryotaro Sahashi, [1]Po-Yen Chen, and [1,2]Teruyasu Mizoguchi
[1]Department of Materials Engineering, The University of Tokyo, Tokyo, Japan
[2]Institute of Industrial Science, The University of Tokyo, Tokyo, Japan

This Supporting Information provides detailed validation results of the machine-learning force-field (MLFF) and Born effective charge (BEC) models, as well as representative structural configurations and supplementary visualization of the electric-field-driven molecular dynamics (MD) simulations.

### S1. Benchmarking of Machine-Learning Force Field and Polarization Models

The predictive performance of the fine-tuned MACE and BEC models described in the Methods was evaluated using independent test data and structural benchmarks.

#### *S1.1. Accuracy of the MACE Force Field*

The MACE model was trained using the publicly available ScAlN dataset reported in Ref.[1], with the training/test split described in the Methods. **Figure S1** shows the correlation between DFT reference values and predictions obtained from the MACE model for the independent test dataset.

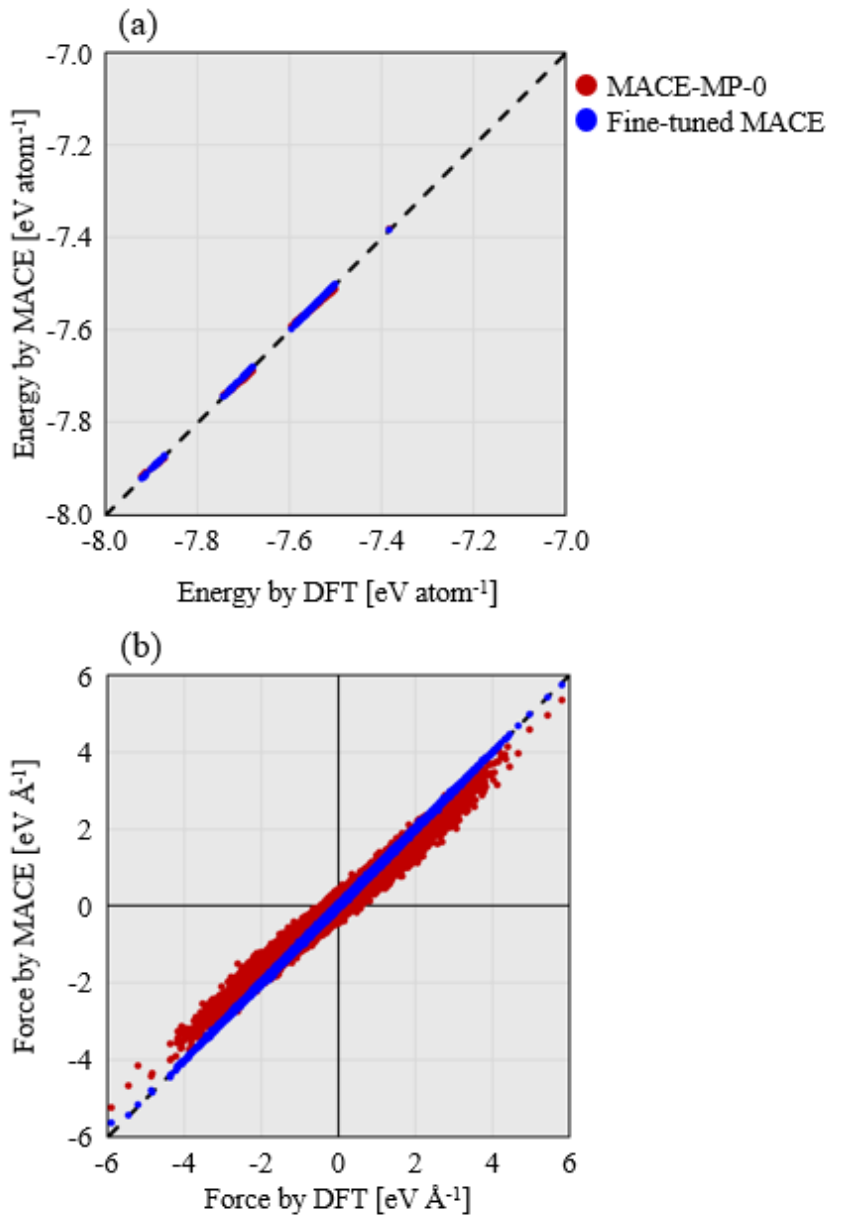

**Figure S1. Correlation between density functional theory (DFT)-calculated values and MACE model predictions for the test dataset.** (a) Total energy (meV atom$^{-1}$) and (b) atomic forces (meV Å$^{-1}$). Red symbols correspond to predictions obtained using the pre-trained MACE model before fine-tuning, while blue symbols represent predictions from the fine-tuned MACE model developed in this study. Results are compared using the same test dataset in both panels. The black dashed lines indicate perfect agreement between the predicted and DFT-calculated values.

As shown in Figure S1(a), which presents the accuracy of total energy predictions, the mean absolute error (MAE) of the pre-trained model (MACE-MP-0-large) was 2.76 meV atom$^{-1}$. After fine-tuning, the MAE was dramatically reduced to 0.22 meV atom$^{-1}$.

Similarly, for atomic force predictions shown in Figure S1(b), the MAE improved substantially from 120.4 meV Å$^{-1}$ before fine-tuning to 6.4 meV Å$^{-1}$ after fine-tuning. Notably, the pre-trained model tends to produce an overly smoothed representation of the potential energy surface, leading to reduced force gradients. Such behavior can artificially soften atomic responses in MD simulations. The fine-tuned model successfully alleviates this limitation, restoring physically meaningful force magnitudes and local gradients that are essential for reliable electric-field-driven MD simulations.

The simultaneous and high-accuracy reproduction of both energies and forces indicates that the constructed model successfully learned the complex PES of ScAlN, which is essential for reliably describing ferroelectric switching dynamics.

To objectively position the performance of the present model, we further conducted a quantitative comparison with other widely used pre-trained general-purpose MLFFs. In addition to MACE-MP-0[2], the following models were considered: SevenNet-0,[3] GRACE-1L-OMAT,[4] ORB v3,[5] and eSEN-30M-OAM.[6] A summary of the prediction accuracies obtained with these models is presented in **Table S1**.

As evident from Table S1, while general-purpose models exhibit reasonable zero-shot performance, their accuracy is insufficient to achieve the sub-meV precision required to describe subtle structural changes and energy barriers associated with ferroelectric behavior.

In contrast, the fine-tuned model developed in this study outperforms these state-of-the-art general-purpose models in terms of both energy and force prediction accuracy, confirming its suitability for high-fidelity MD simulations of ferroelectric switching in ScAlN.

**Table S1. Prediction accuracy of different machine-learning force field models.** Energy and force mean absolute errors (MAEs) are compared among different machine-learning force field (MLFF) models. The fine-tuned MACE model developed in this study achieves sub-meV energy accuracy and outperforms the universal MLFFs examined here.

| MLFF Model | Energy MAE / meV atom$^{-1}$ | Force MAE / meV Å$^{-1}$ |
|---|---|---|
| Fine-tuned MACE | 0.22 | 6.42 |
| MACE-MP-0 | 2.76 | 120.40 |
| SevenNet-0 | 3.63 | 75.41 |
| GRACE-1L-OMAT | 45.52 | 82.04 |
| ORB v3 | 14.35 | 22.77 |
| eSEN-30M-OAM | 1.12 | 14.62 |

*S1.2. Reproducibility of Structural Parameters*

The reproducibility of crystal structures, which is directly linked to ferroelectric properties, was also carefully examined.

In wurtzite-type ferroelectrics, the *c*/*a* ratio is a key structural parameter that governs both the spontaneous polarization and the height of the switching barrier. Therefore, accurately capturing its composition dependence is critically important.[7]

We evaluated the reproducibility of the lattice parameters—namely, the *a*-axis length, the *c*-axis length, and the resulting *c*/*a* ratio—as summarized in **Figure S2**(a).

The fine-tuned model successfully reproduced the experimentally established trend that the *c*/*a* ratio decreases with increasing Sc concentration. Quantitative agreement with experimental values was obtained across the entire composition range from $x = 0$ to $x = 0.375$.

Because the *c*/*a* ratio plays a decisive role in determining spontaneous polarization and switching barriers in wurtzite ferroelectrics, this high level of structural accuracy provides a solid foundation for the reliability of the MD simulations discussed in subsequent sections.

*S1.3. Accuracy of the Born Effective Charge Model*

To assess the accuracy of the BEC model described in the Methods, **Figure S3** compares the predicted BECs with the density functional perturbation theory (DFPT) reference values for Sc, Al, and N atoms. When the initial BM1 model prior to fine-tuning was used, large mean absolute errors (MAEs) were observed for all atomic species (Sc: 2.744 e, Al: 2.322 e, N: 3.033 e). This poor performance is attributed to the absence of wurtzite-type structures in the training data of the initial model.

In contrast, after fine-tuning using the ScAlN-specific dataset, the MAEs were dramatically reduced to 0.020 e for Sc, 0.011 e for Al, and 0.014 e for N. The fine-tuned model thus achieved highly accurate reproduction of DFPT-calculated BEC values.

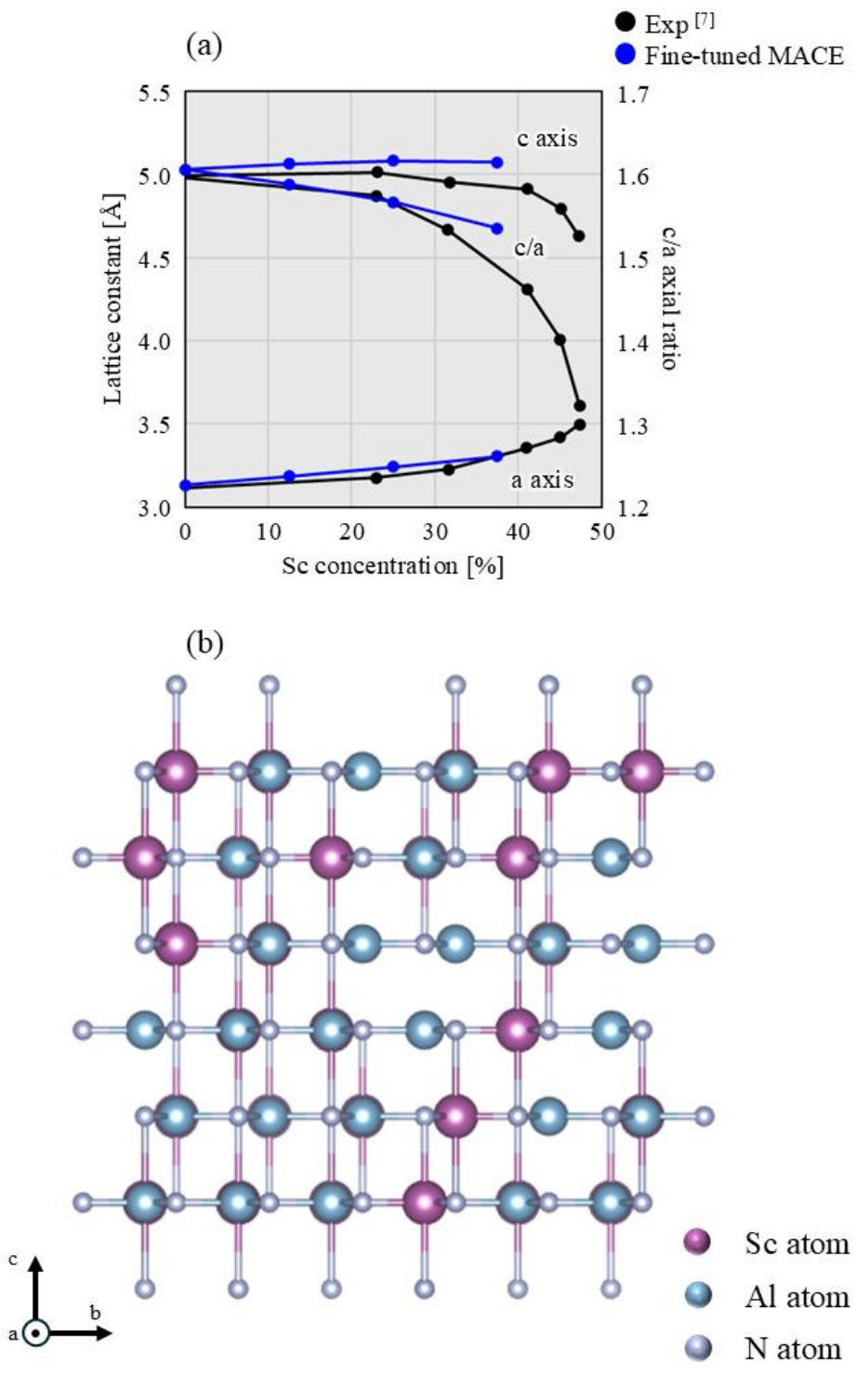

**Figure S2. Validation of structural properties and definition of the planar hexagonal reference structure in ScAlN.** (a) Composition dependence of crystal structural parameters as a function of Sc concentration. Lattice constants *a* and *c* are shown with reference to the left vertical axis, while the *c*/*a* ratio is shown with reference to the right vertical axis within a single plot. Blue symbols represent results calculated using the fine-tuned MACE model, and black symbols indicate experimental values taken from Ref.[13]. (b) Schematic illustration of the composition-specific planar hexagonal nonpolar reference used for polarization evaluation in this study. For each composition, the Sc/Al occupations, in-plane atomic coordinates, and simulation cell are retained, while each N atom is positioned in the nearest cation plane along the *c* direction. This configuration represents the planar midpoint of the wurtzite switching coordinate and is used as the reference for both the species-resolved displacement and relative-polarization analyses. Purple, blue, and light-violet spheres represent Sc, Al, and N atoms, respectively.

These results confirm that the electric-field-induced driving forces acting on individual atoms are described with first-principles-level accuracy, which is essential for reliable electric-field-induced MD simulations.

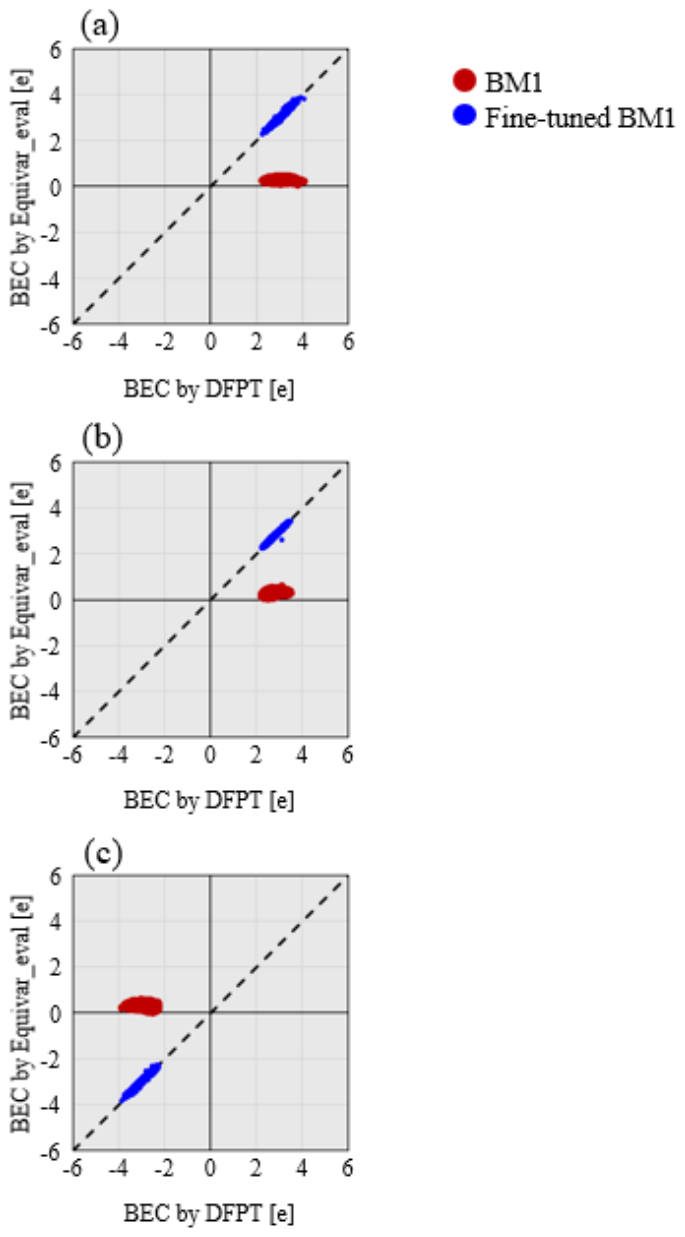

**Figure S3. Prediction accuracy of Born effective charges (BECs).** Comparisons between density functional perturbation theory (DFPT)-calculated values and equivar_eval model predictions for (a) Sc, (b) Al, and (c) N atoms. Red symbols correspond to predictions from the BM1 model before fine-tuning, while blue symbols represent predictions obtained after fine-tuning. The black dashed lines indicate perfect agreement between the predicted and DFPT-calculated values.

## S2. Validation of Simulation Setup

The simulation conditions described in the Methods were quantitatively validated using the representative composition $Sc_{0.25}Al_{0.75}N$. The following tests evaluate the robustness of the calculated ferroelectric properties against the reference structure, interaction range, supercell size, electric-field application scheme, and sweep rate.

### *S2.1. Reference Structure for Polarization Evaluation*

The composition-specific planar hexagonal nonpolar reference defined in the Methods is illustrated in Figure S2(b). In this configuration, each N atom lies in the nearest cation plane along the *c* direction. This structure represents the planar midpoint of the wurtzite switching coordinate and was used consistently as the reference for both the polarization and species-resolved displacement analyses. Because the random Sc/Al occupation in the finite alloy supercell does not necessarily preserve exact inversion symmetry, this configuration is referred to as a planar hexagonal nonpolar reference rather than a strictly centrosymmetric structure.

### *S2.2. Interaction Range: Short-Range vs Long-Range Model*

**Figure S4**(a) compares the default "Short Range" MACE model with a long-range interaction model based on a locally extended scheme (LES), referred to as the "Long Range" model[8], using a 2880-atom supercell. The two curves almost completely overlap, indicating that the interaction cutoff treatment does not significantly influence the ferroelectric switching behavior.

Quantitatively, coercive field ($E_c$) and remanent polarization ($P_r$) obtained using the default Short Range model were 8.82 MV cm$^{-1}$ and 107 μC cm$^{-2}$, respectively. These values are in excellent agreement with those obtained using the Long Range model, namely 9.44 MV cm$^{-1}$ for $E_c$ and 106 μC cm$^{-2}$ for $P_r$, with only negligible differences.

Based on these results, we adopted the default Short Range model in this study, as it provides an optimal balance between computational efficiency and accuracy.

### *S2.3. Validation of the Supercell Size*

Next, to assess the impact of finite-size effects, we compared simulations performed using a standard 360-atom supercell and a larger 2880-atom supercell (Figure S4(b)).

A comparison of the calculated ferroelectric properties shows that the 360-atom system yields $E_c$ = 9.33 MV cm$^{-1}$ and $P_r$ = 106 μC cm$^{-2}$, whereas the corresponding values for the 2880-atom system are $E_c$ = 8.82 MV cm$^{-1}$ and $P_r$ = 107 μC cm$^{-2}$. The close agreement between these values indicates that the results are sufficiently converged with respect to finite-size effects even for the 360-atom supercell. Therefore, we conclude that the 360-atom system is large enough to represent bulk ferroelectric properties, and this supercell size was adopted for all subsequent analyses.

It should be noted that, in the above validation tests, the ferroelectric properties were evaluated under quasi-static conditions by varying the applied electric field stepwise in increments of 1 MV cm$^{-1}$. In contrast, the switching dynamics analyses presented in the following sections employed a continuous electric-field ramp in order to track atomic motions continuously. Under the stepwise protocol, the $E_c$ = 9.33 MV cm$^{-1}$ and $P_r$ = 106 μC cm$^{-2}$, whereas under the continuous-ramp protocol, $E_c$ = 10.12 MV cm$^{-1}$ and $P_r$ = 107 μC cm$^{-2}$.

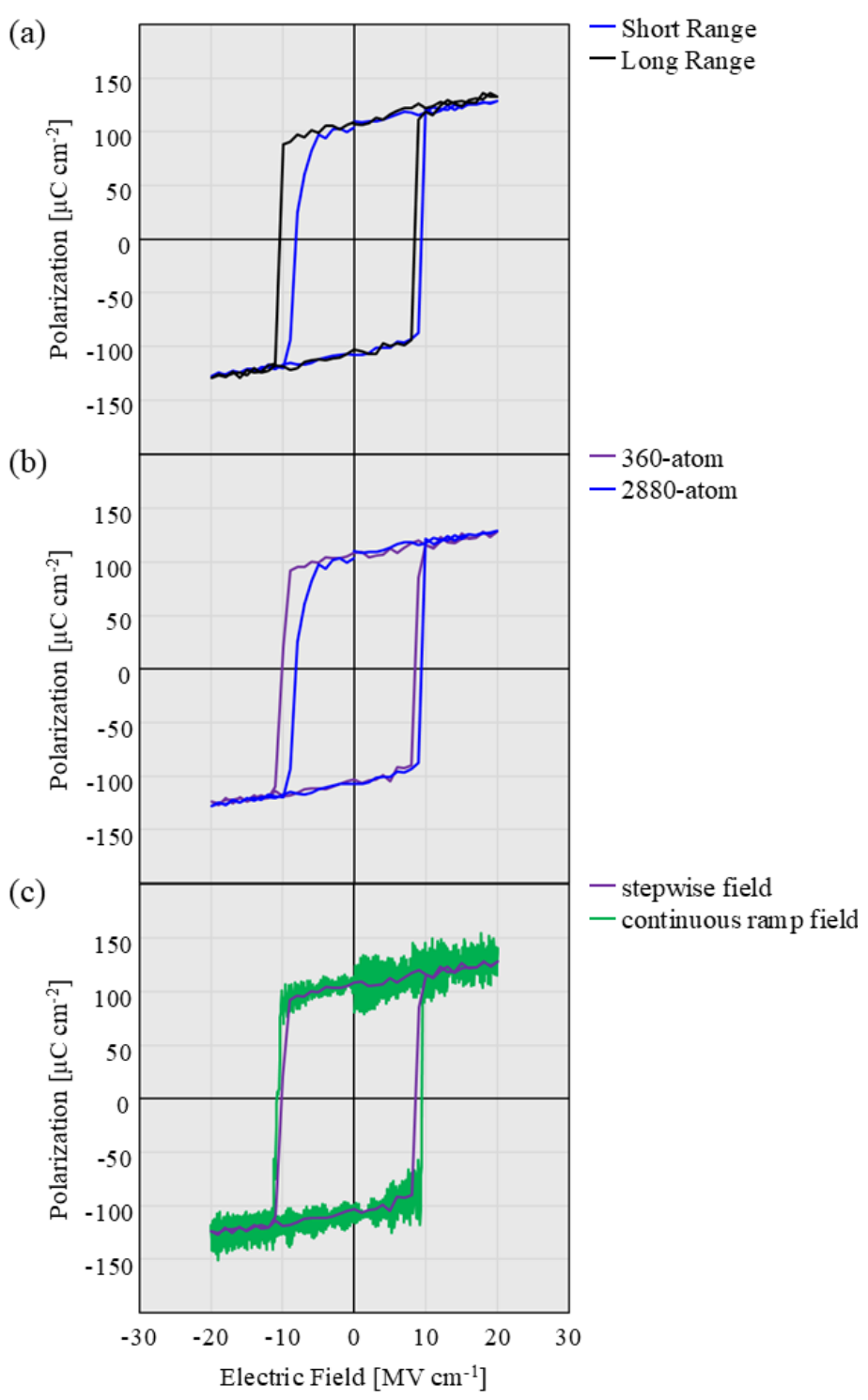

**Figure S4. Validation of the simulation conditions used for electric-field-induced molecular dynamics simulations.** (a) Comparison of interaction range effects. *Polarization-electric field* (*P-E*) hysteresis loops obtained for a 2880-atom supercell. Blue curves correspond to the default Short Range model, and black curves correspond to the locally extended scheme (LES)-type Long Range model. (b) Comparison of supercell size dependence. *P-E* hysteresis loops obtained using a 360-atom supercell (purple) and a 2880-atom supercell (blue). (c) Comparison of *P-E* hysteresis loops obtained using different electric-field application schemes. Results obtained using the stepwise field application are shown in purple, while those obtained using the continuous ramp field application are shown in green.

These results are also evident from the *P-E* hysteresis loops shown in Figure S4(c), which demonstrate that the overall loop shape and the key ferroelectric parameters remain essentially unchanged despite the different electric-field application schemes. This confirms that the choice of electric-field protocol does not significantly affect the primary ferroelectric properties. Although the specific electric-field application methods differ, the underlying MD framework is identical; therefore, the absence of significant differences in the ferroelectric characteristics is consistent with expectations.

The real-time atomic evolution during electric-field-driven switching under the continuous ramp protocol is shown in **Movie S1**, illustrating the simulation framework adopted in this study.

Taken together, these validations confirm that the present simulation setup — including reference structure definition, interaction cutoff choice, supercell size, and electric-field sweep protocol — provides a quantitatively reliable and converged description of ferroelectric switching in ScAlN.

**Movie S1. Real-time atomic evolution during electric-field-induced polarization switching in $Sc_{0.25}Al_{0.75}N$.** The movie was obtained from machine-learning force-field (MLFF)-based molecular dynamics (MD) simulations under a continuous electric-field ramp of 0.05 kV cm$^{-1}$ fs$^{-1}$ and illustrates the large-scale simulation framework employed in this study.

### *S2.4. Dynamic response and sweep-rate dependence*

The electric-field sweep rate is a critical parameter in ferroelectric switching simulations because $E_c$ is not an intrinsic constant but a time-dependent observable determined under finite switching time conditions. Experimentally, the $E_c$ of ScAlN increases systematically with measurement frequency, reflecting the exponential time dependence of domain nucleation and growth described by Merz's law.[9] According to this relationship, the characteristic switching time decreases exponentially with increasing electric field. Consequently, when the available switching time becomes shorter at higher frequencies (or faster sweep rates), a larger electric field is required to complete polarization reversal within the given timescale. This strong rate dependence is particularly pronounced in ScAlN due to its unusually high activation field. Reproducing this experimentally observed behavior in simulations therefore serves as a stringent benchmark for validating not only the accuracy of the interatomic forces and BECs, but also the physical fidelity of the electric-field-driven MD protocol itself.

Using the constructed models, we analyzed polarization switching behavior under three different electric-field sweep rates. **Figure S5**

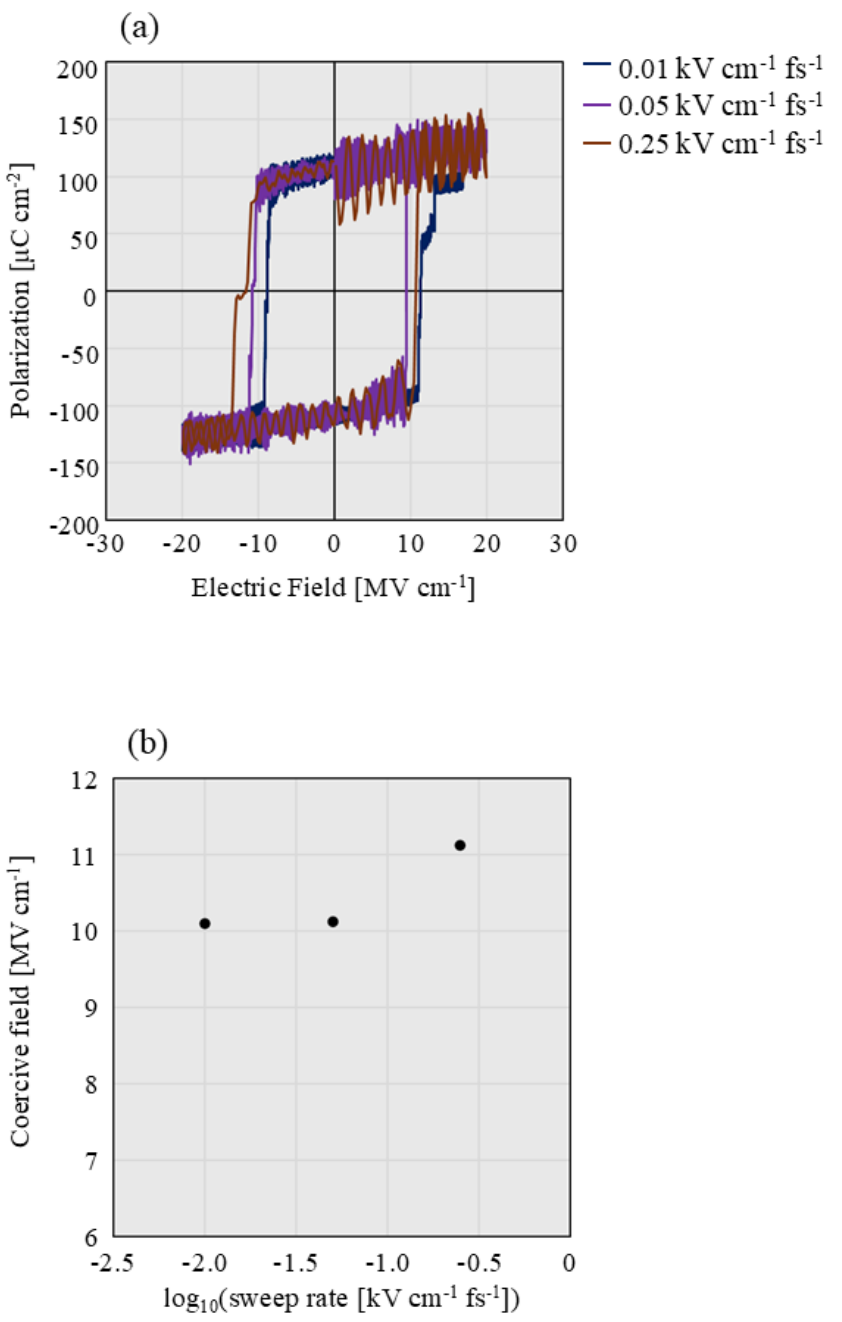

**Figure S5. Dependence on electric-field sweep rate.** (a) *Polarization-electric field* (*P*-*E*) hysteresis loops obtained at sweep rates of 0.01 kV $cm^{-1}$ $fs^{-1}$ (navy), 0.05 kV $cm^{-1}$ $fs^{-1}$ (purple), and 0.25 kV $cm^{-1}$ $fs^{-1}$ (brown). (b) Coercive field ($E_c$) as a function of sweep rate. The horizontal axis is shown on a base-10 logarithmic scale. Black points represent the calculated $E_c$ values.

summarizes the sweep-rate-dependent switching characteristics. Figure S5(a) shows the *P*-*E* hysteresis loops obtained at each sweep rate.

As a general trend, increasing the electric-field sweep rate leads to a wider hysteresis loop, indicating that a higher electric field is required to induce polarization reversal under faster field variations.

The $E_c$ extracted from each hysteresis loop is plotted as a function of the electric-field sweep rate in Figure S5(b), where the horizontal axis is shown on a logarithmic scale.

Although the applied electric-field frequency is significantly higher than the Hz-kHz range typically used in experimental *P*-*E* measurements, this choice is dictated by the intrinsic time-scale limitations of MD simulations. In MD, the time step is restricted to the order of femtoseconds in order to resolve the fastest atomic vibrations. As a consequence, the accessible simulation time is typically limited to nanoseconds, which leads to an intrinsic "timescale gap" between atomistic simulations and macroscopic experiments[10]. Therefore, reproducing experimental Hz-kHz frequencies within a fully atomistic MD framework is computationally infeasible and conceptually unnecessary.

At the highest sweep rate of 0.25 kV $cm^{-1}$ $fs^{-1}$, a relatively large $E_c$ = 11.13 MV $cm^{-1}$ was obtained. In contrast, when the sweep rate was reduced to 0.05 kV $cm^{-1}$ $fs^{-1}$ and further to 0.01 kV $cm^{-1}$ $fs^{-1}$, the $E_c$ decreased to 10.12 MV $cm^{-1}$ and 10.09 MV $cm^{-1}$, respectively. The difference between these two lower sweep rates is as small as approximately 0.03 MV $cm^{-1}$.

These results suggest two important points. First, the present simulation framework successfully reproduces the characteristic dynamic rate dependence of ferroelectric switching, in which polarization reversal is delayed under rapidly varying electric fields.[11] Second, at a sweep rate of 0.05 kV $cm^{-1}$ $fs^{-1}$, the $E_c$ has already converged to a value close to the quasi-static limit, and further reduction of the sweep rate does not lead to any significant change in the results.

Therefore, considering the balance between computational cost and accuracy, we identified 0.05 kV $cm^{-1}$ $fs^{-1}$ as the optimal electric-field sweep rate. This value provides sufficient convergence while retaining physical fidelity, and it was adopted for all subsequent analyses.

## S3. Evaluation of Configuration Dependency on Ferroelectric Properties

To evaluate the influence of local atomic configuration randomness on the ferroelectric properties, we performed identical electric-field-induced MD simulations on an alternative SQS configuration (Configuration 2) and compared it with the primary configuration (Configuration 1) used in the main text. The randomness of the cation configurations was quantified using the SQS pair-correlation analysis described in the Methods, following Ref.[12].

We evaluated Configurations 1 and 2 for each composition. The maximum absolute errors among the first to sixth nearest-neighbor shells for each structure are as follows:

At $x$ = 0.125: 0.0069 for both Configuration 1 and Configuration 2.

At $x$ = 0.25: 0.0056 for both Configuration 1 and Configuration 2.

At $x$ = 0.375: 0.0069 for Configuration 1 and 0.0116 for Configuration 2.

Furthermore, the mean absolute errors across the six neighbor shells ranged from 0.0016 to 0.0049. Therefore, none of the generated atomic configurations exhibit significant Sc-Al short-range ordering up to at least the sixth nearest-neighbor shell, confirming that they accurately reproduce the pair correlations of an ideal random solid solution.

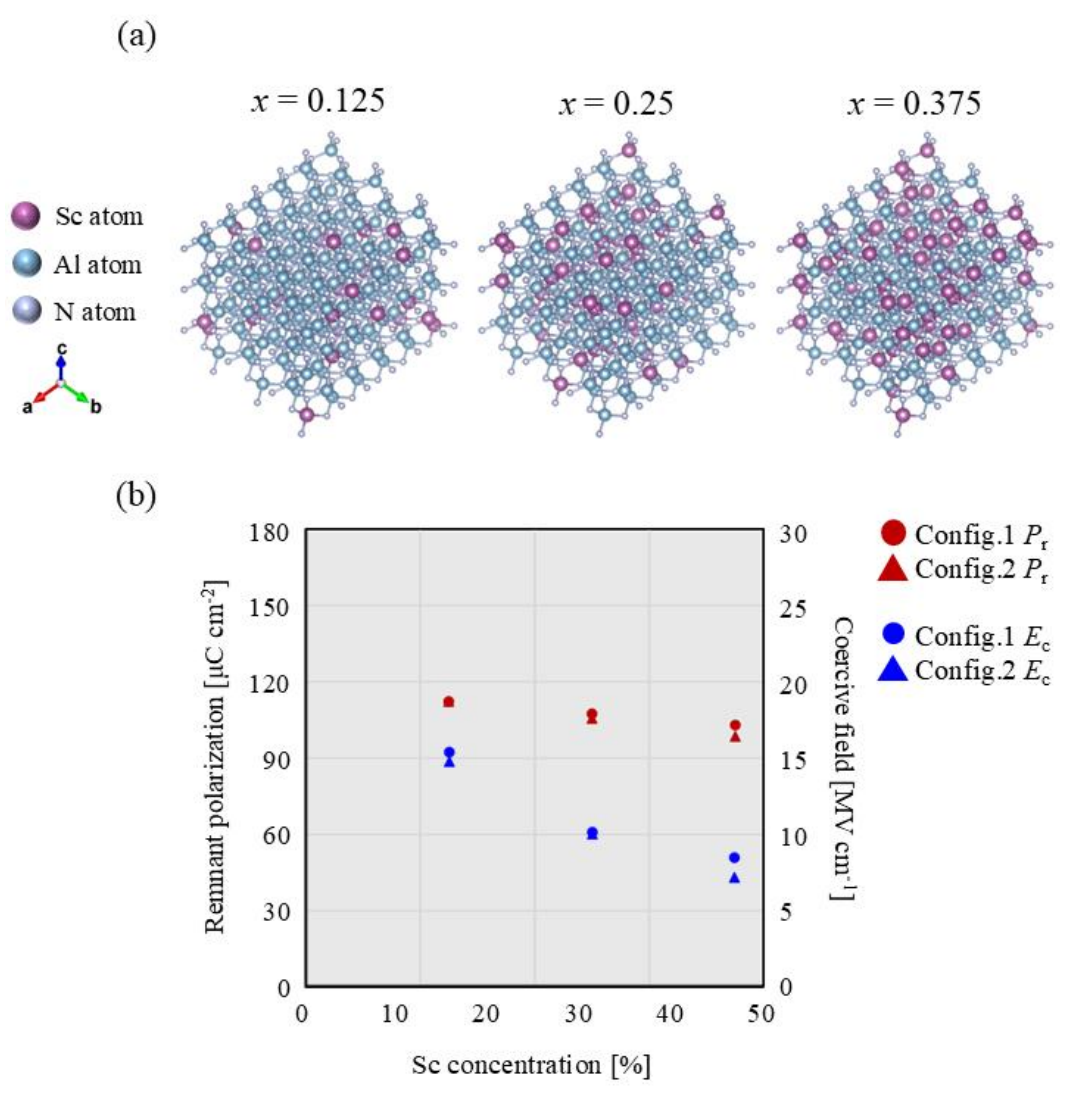

**Figure S6. Effect of special quasirandom structure configurations on the calculated ferroelectric properties.** (a) Atomistic model visualizations of the primary special quasirandom structure (SQS) supercells (Configuration 1) for $x$ = 0.125, 0.25, and 0.375, visually illustrating the random distribution of Sc atoms. Purple, blue, and light-violet spheres represent Sc, Al, and N atoms, respectively. (b) Composition dependence of the $P_r$ (left axis) and $E_c$ (right axis) for the primary configuration used in the main text (circles) and an alternative SQS configuration (triangles). The vertical axis on the left represents $P_r$, and the vertical axis on the right represents $E_c$. Red symbols represent the remanent polarization ($P_r$), whereas blue symbols represent the coercive field ($E_c$); circles and triangles denote Configurations 1 and 2, respectively.

As shown in **Figure S6**(a), the constructed SQS supercells exhibit a visually uniform and random distribution of Sc atoms without any artificial macroscopic clustering. Furthermore, Figure S6(b) demonstrates that while minor variations exist between the configurations, the difference in $E_c$ at the same composition is approximately 1.3 MV cm$^{-1}$ (at $x = 0.375$). This difference is significantly smaller than the macroscopic composition-dependent decrease in $E_c$ (e.g., a substantial reduction of over 5 MV cm$^{-1}$ from $x = 0.125$ to $x = 0.25$). Similarly, the variations in $P_r$ are small and do not obscure the overall decreasing trend with increasing Sc content. These results clearly demonstrate that the pronounced composition dependence of the switching characteristics is highly robust. The local atomic configuration does not alter the fundamental physical mechanism or the macroscopic trends discussed in the main text, verifying that extensive high-throughput structural screening is not required to capture the intrinsic switching behavior in this MLFF-based MD framework.

## S4. Statistical Summary of the Solid-Angle Analysis

To complement the solid-angle distributions shown in Figure 5(d-f) of the main text, we calculated the mean and standard deviation of the N-view solid angle for both the actual and virtual trajectories at each Sc concentration. The results are summarized in **Table S2**. For the actual trajectories, the mean solid angle increases modestly from 3.418 sr at $x = 0.125$ to 3.475 sr at $x = 0.375$. These statistical values provide a quantitative summary of the overall solid-angle distributions, while the detailed distribution profiles and their characteristic features are discussed in the main text.

**Table S2. Statistical summary of the N-view solid angle.** Mean and standard deviation of the solid angle of the cation triangle viewed from the N atom for the actual and virtual trajectories at each Sc concentration are listed. Solid angles are reported in steradians (sr).

| Sc concentration $x$ | Actual (sr) | Virtual (sr) |
|---|---|---|
| 0.125 | 3.418 ± 0.365 | 3.541 ± 0.311 |
| 0.250 | 3.423 ± 0.384 | 3.370 ± 0.390 |
| 0.375 | 3.475 ± 0.430 | 3.356 ± 0.440 |